# Backward Mapping from Device Targets to Chemical Genomes for Interpretable Discovery of Phase-Stable Lead-Free Double Perovskites with DFT-Validated Design Rules

Nafis Ahtasum [1,2,*], Sohanur Rahman Sohan[1,3], Md. Mostaq Ahmed Himel[1,2], Md. Zahid Hassan[3], Muhammad Harussani Moklis[1,4,5], Md Rafiul Alam Roni[6]

[1]Center for Material, Climate and Energy, Research and Analysis Institute, RAI Initiative Ltd, Dhaka, Bangladesh

[2]Department of Apparel Engineering, Bangladesh University of Textiles, Dhaka-1208, Bangladesh

[3]Department of Textile Engineering Management, Bangladesh University of Textiles, Dhaka-1208, Bangladesh

[4]Department of Chemical and Environmental Engineering, Faculty of Engineering, Universiti Putra Malaysia, Serdang 43400, Selangor, Malaysia

[5]Energy Science and Engineering, Department of Transdisciplinary Science and Engineering, Institute of Science Tokyo, 2-12-1, Ookayama, Meguro-ku, Tokyo 152-8550, Japan

[6]Department of Chemical Engineering, University of Liege, Liege 4000, Belgium

*=corresponding author (email= nafisahtasum666@gmail.com)

## Abstract

Lead-free halide double perovskites are promising alternatives to Pb-based semiconductors, but their discovery is constrained by the need to simultaneously satisfy structural formability, thermodynamic stability, band-gap placement, optical-transition strength, dielectric screening, and carrier transport within a vast $A_2BB'X_6$ chemical space. Here, we present a backward-mapping, genome-guided framework that links device-level targets to chemically interpretable descriptor families for Pb-free double-perovskite discovery. Starting from 13,088 charge-balanced compositions, we implement a halide-aware screening workflow integrating geometric formability filtering, six-family chemical-genome descriptor encoding, evolutionary-optimized machine-learning surrogates, SHAP-based interpretation, and DFT phenotype closure. The stability model is trained using $E_{hull}$-derived labels, while the band-gap surrogate predicts scalar-relativistic PBE $E_g$ for target-driven selection. This staged inverse-design funnel reduces the search space to seven DFT-validated candidates: $K_2BePdF_6$, $K_2MnCdCl_6$, $Rb_2TeCuBr_6$, $Cs_2SnGeBr_6$, $Cs_2GeSrBr_6$, $Cs_2NiBaI_6$, and $Cs_2AgInCl_6$. All structural assignability, band-edge orbital character, effective masses, dielectric response, optical absorption, conductivity, reflectivity, energy-loss spectra and XRD fingerprints have been verified by first-principles calculation. For RbTeCuBr and CsSnGeBr, their absorber-like characteristics are well-balanced. CsGeSrBr and CsAgInCl are preferable in the UV-active and electron-selective role respectively. Beyond candidate identification, the framework extracts DFT-validated design rules from six chemical-gene families: packing, cohesion, polarity/covalency, charge transfer, polarization, and electronic identity; revealing that functional materials emerge from a coupled stability–function intersection rather than band-gap optimization alone. This work establishes an interpretable inverse-design paradigm for navigating Pb-free double-perovskite space and accelerating discovery of optoelectronic materials.

## 1. Introduction

Halide perovskites are a class of semiconducting materials. It has some extraordinary properties such as efficient light absorption and adjustable bandgap. They have long carrier diffusion lengths and superior defect tolerance with low-temperature processability. They show several favorable properties. These include strong light absorption, tunable band gaps, and good charge transport behavior. Because of these, they are widely studied for solar cells, photodetectors, light-emitting diodes, radiation detectors, and scintillators.

Almost all of the top performance halide perovskites contain Pb. This is problematic because Pb is poisonous and pollutes the surrounding environment. The compounds do not have high stability and they can degrade in the presence of water. They can also undergo thermal decomposition and ion migration. They do not always show good long-term stability. The discovery of Pb-free perovskites is an important scientific task. [1]

In recent years, halide double perovskites with the chemical composition $A_2BB'X_6$ gained substantial attention as promising alternatives to Pb-based perovskites. The main feature of halide double perovskites is the replacement of $Pb^{2+}$ cation in typical $ABX_3$ perovskites by two metal cations organized in $B^{+}/B'^{3+}$ or $B^{2+}/B'^{2+}$ configurations. [2,3] Such a substitution preserves the crystal structure characteristic of conventional perovskites and expands the set of possible compounds.

The properties of halide double perovskites, including structural stability, thermodynamic stability, bandgap, optical absorption, dielectric response, and charge carrier transport can be controlled by varying the cations at the A and B sites and X halides. Consequently, Pb-free double perovskites can be successfully employed as materials for solar cells, photodetectors, transparent electronics, ultraviolet sensors, scintillators, and radiation detectors.

Currently, high-throughput density functional theory and machine learning techniques prove to be effective approaches in searching for novel Pb-free perovskites.[4] DFT provides reliable material properties but becomes computationally prohibitive for extensive chemical spaces, whereas ML can predict properties rapidly after training on DFT data. [5] Previous studies have combined ML and DFT to predict band gap, formation energy, stability, and related properties of halide perovskites. Landini trained a neural-network model for band-gap prediction and screened 7,056 lead-free halide double perovskites, followed by hybrid-DFT evaluation of a reduced subset.[6] Chen developed ML models using a dataset of 3,720 $ABX_3$ perovskites and 2,660 double perovskites, targeting both band gap and formation energy while emphasizing interpretability through feature-attribution analysis.[7] Gao performed high-throughput DFT screening of 760 $Cs_2B^{2+}B'^{2+}X_6$ compositions for lead-free photovoltaic absorbers.[8]

From these examples, it can be seen that computation provides a powerful way to speed up discovery of Pb-free perovskites. However, there are major drawbacks. Most prior works have used forward screening, in which a list of compositions is first generated, properties are then calculated and the compositions selected on the basis of those properties. It is unclear what a connection is between the properties of the material and the requirement for the device. Also, many works investigate one or two properties only, such as the bandgap or formation energy. In order for a material to work in a device it needs to be phase stable, have a suitable bandgap and absorb light well, have sufficient dielectric screening and allow charge transport. Single-property optimization is therefore insufficient for reliable materials discovery.

Pb-free double perovskites exhibit complex property interrelationships. $Cs_2AgBiBr_6$, a well-known example, is more stable than many Pb-based perovskites, yet its performance is limited by a wide and often indirect band gap, weak absorption near the band edge, and modest carrier transport. Similar trade-offs appear broadly: some materials are stable but optically poor, others exhibit favorable band gaps but low carrier mobility, and still others show strong dielectric response but weak absorption. The central challenge is not merely identifying Pb-free materials, but finding those where multiple properties are balanced simultaneously.

Lead-free double-perovskite discovery is therefore a strongly multi-objective problem. Charge neutrality generates many nominally valid formulas, but only a small fraction is expected to be geometrically formable and functionally useful. Ionic-size mismatch can prevent formation of the corner-sharing octahedral framework. Weak or unbalanced metal–halide bonding reduces thermodynamic stability. Localized d-derived band edges produce flat bands and heavy carriers. Heavy halides may narrow the band gap and increase polarizability, but they can also shift the structure outside the optimal packing window. These trade-offs demonstrate that useful candidates cannot be reliably identified from a single descriptor, property, or ranking score. A more effective strategy must connect device-level requirements to chemically interpretable variables amenable to rational tuning.[9,10]

Interpretability is particularly important because different material properties are governed by distinct chemical factors. Structural formability is primarily controlled by ionic radii and tolerance-related packing descriptors. Thermodynamic stability depends on framework cohesion, metal–halide bonding, and competing decomposition pathways. Band-gap tuning is strongly influenced by B/B′-site orbital identity, electronegativity contrast, and metal–halide hybridization. Dielectric and optical responses depend on polarizability, band-edge transition strength, and electronic screening. Carrier transport is governed by band dispersion and the orbital character of valence- and conduction-band edges[10–13]. A useful discovery framework must therefore not only predict which compositions may succeed but also explain why they succeed and how related chemistries can be improved.

In this work, we develop a backward-mapping genome guided framework to connect chemical genome (descriptor) families for discovery of Pb-free $A_2BB'X_6$ double perovskites with device-level targets (Figure 1). Unlike conventional forward prediction from composition to property, this approach begins with device requirements and maps backward to chemical descriptor "genes," treating DFT-calculated properties as observable phenotypes. Starting from 13,088 charge-balanced Pb-free compositions, we apply geometric formability filters and encode each composition using six interpretable chemical-genome families: geometric packing, framework cohesion, polarity/covalency, charge transfer, polarization/screening, and electronic identity. Evolutionary-optimized ML surrogate models, trained on an independently assembled DFT-labeled dataset, predict thermodynamic stability and PBE band gap, with the stability model optimized for high recall to avoid premature elimination of promising candidates. SHAP-based interpretation decodes the learned model behavior into chemically meaningful design rules, while a staged inverse-design funnel followed by DFT structural, electronic, optical, and transport validation compresses the broad chemical space to a small set of actionable candidates. This work advances Pb-free double-perovskite discovery from broad forward enumeration toward a backward-mapped, chemically explainable inverse-design strategy.

## 2. Methods

### 2.1 Backward-mapping discovery workflow

The methodological framework was designed as a backward-mapping discovery workflow in which device-relevant requirements are translated into chemically interpretable screening criteria and then projected back onto the lead-free composition space[14]. Rather than treating inverse design as a purely algorithmic ranking problem, the workflow integrates exhaustive chemical-space enumeration, halide-aware formability filtering, interpretable descriptor engineering, machine-learning surrogate modeling, and first-principles validation. This design balances three essential requirements for high-throughput materials discovery: broad chemical coverage, physical interpretability, and DFT-level verification[15,16].

The workflow comprises five coupled stages. First, a charge-balanced Pb-free library is constructed across heterovalent and homovalent oxidation-state families. Second, geometry-based formability filters remove structurally implausible candidates. Third, each composition is encoded using a six-family chemical-genome descriptor framework. Fourth, machine-learning surrogate models are trained using an independently assembled DFT-labeled reference dataset to predict thermodynamic stability and scalar-relativistic PBE band gap. Finally, target-driven screening is followed by DFT phenotype closure of prioritized candidates, including structural relaxation, electronic-structure analysis, effective-mass extraction, optical-response calculations, and simulated XRD fingerprinting.

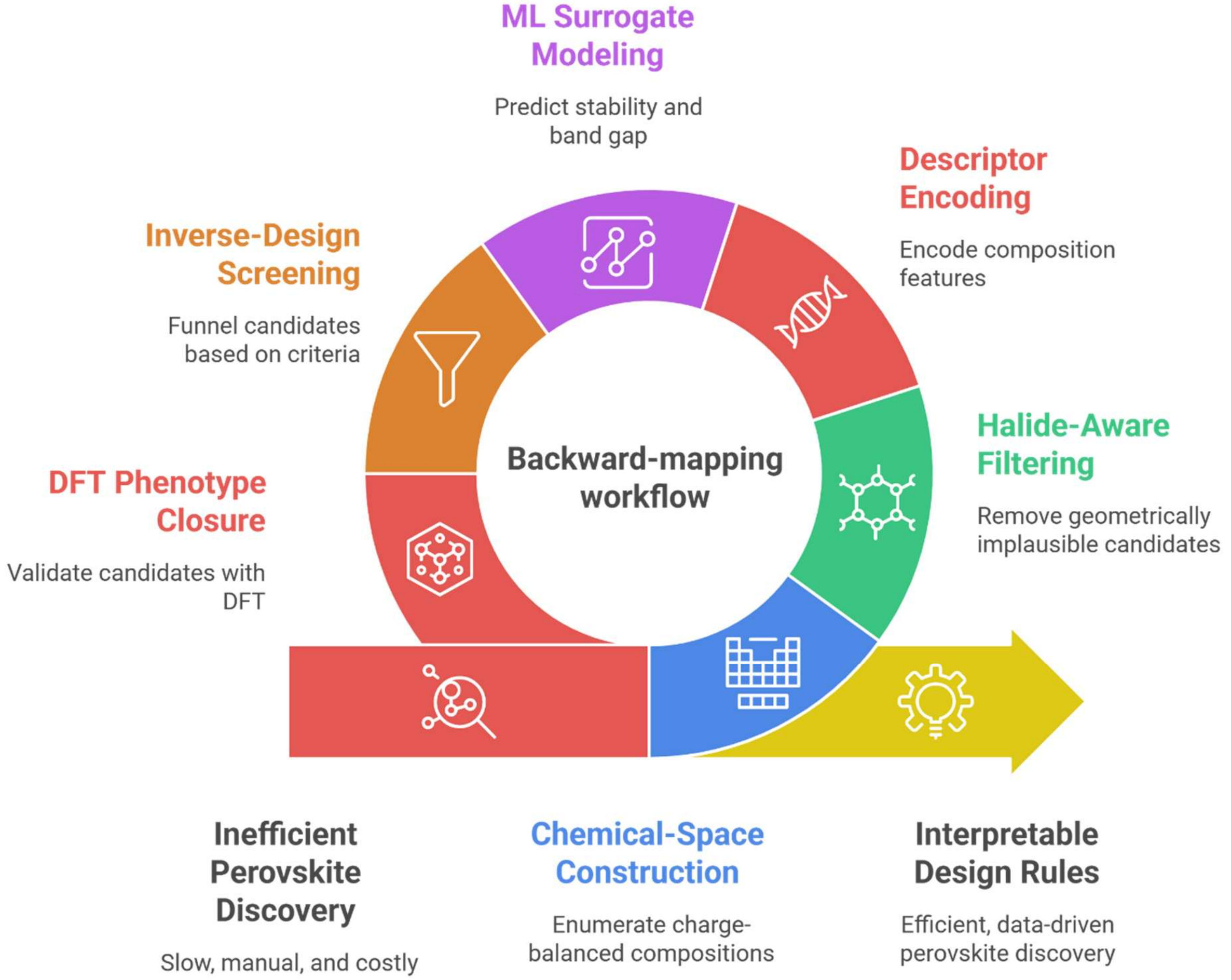

Figure 1. Genome-guided backward-mapping workflow used in this study. The workflow combines charge-balanced chemical-space enumeration, six-family chemical-genome descriptor construction, evolutionary-optimized machine-learning surrogate models, staged inverse-design screening, and DFT phenotype closure for structural, electronic, transport, optical, and XRD-based validation.

2.2 Construction of the $A_2BB'X_6$chemical space

A Pb-free all-inorganic double-perovskite chemical space was generated following the generic formula AB B X, where A is a mono-valent cation, and $B/B'$ are metal-site cations and X a halogen anion. Two charge-neutral oxidation state categories of compounds were explored: heterovalent and homovalent (where $B/B'$corresponds to mixed or identical oxidation states for the two metal sites).The available monovalent cations were the Alkali and Alkali metal groups (Li to Cs), the copper, silver and gold metals and the halide-site was kept for F, Cl, Br and I. The available A-site cations were from alkali metals Li-Cs, some transition metals and the group of semi-metals including transition metals like Pd, Ir and Pt and post-transition metals like Al-Bi.

All possible compounds were generated through a Python work-flow implemented in the pymatgen library. A composition is retained if and only if the following nominal charge neutrality condition holds: $2q_A + q_B + q_{B'} + 6q_X = 0$,
where $q_A$, $q_B$, $q_{B'}$, and $q_X$ are the oxidation states of the individual ions involved.

For the homovalent family, $B/B'$pairs were canonicalized to remove duplicate formulas caused by site permutation. This enumeration produced 13,088 unique charge-balanced Pb-free compositions, which formed the starting library for the screening workflow[17].

### 2.3 Halide-aware geometric formability filtering

Geometry-based formability descriptors were calculated for each enumerated composition using Shannon ionic radii. The Goldschmidt tolerance factor t, octahedral factor μ, and new tolerance factor τ were employed as first-pass structural feasibility descriptors. The accepted formability windows were set as $0.8 \leq t \leq 1.1$, $\mu \geq 0.41$, and $\tau < 4.18$ [18–20]. These criteria help identify compositions that are geometrically compatible with a corner-sharing double-perovskite framework. The same quantities were retained as interpretable geometric packing genes in the chemical-genome descriptor set, maintaining connectivity between the formability filter and subsequent model-interpretation and design-rule analysis.

### 2.4 DFT-labeled reference dataset and learning targets

Machine-learning models were trained using a DFT-labeled reference dataset of 1,221 unique halide double perovskites assembled through a reproducible Python workflow based on Materials Project data. The dataset included 401 homovalent and 820 heterovalent compounds, allowing the models to learn from both oxidation-state families.

Thermodynamic stability was quantified using the energy above the convex hull, $E_{hull}$, as a 0 K phase-stability descriptor. Compounds satisfying $E_{hull} \leq 25$ meV/atom were labeled as stable (1), while compounds above this threshold were labeled as unstable (0)[21]. The scalar-relativistic PBE band gap, $E_g$, served as the continuous regression target. Data was partitioned into training and test sets with an 80/20 stratified split according to the stability label. [9]The E based label reflects DFT calculated thermodynamic stability at 0K and is independent of finite temperature effects, anharmonic lattice dynamics, kinetic trapping, sensitivity to moisture/oxygen or long-term degradation.

The term "stable" in the machine-learning stage therefore refers to DFT-predicted thermodynamic stability rather than guaranteed experimental stability under ambient conditions.

Table 1. DFT-labeled dataset and learning-target definitions.

| Item | Description |
| --- | --- |
| System | $A_2BB'X_6$halide double perovskites[11] |

| Dataset size | 1,221 compounds |
| --- | --- |
| Data source | Materials Project-derived DFT data |
| Composition families | 401 homovalent; 820 heterovalent |
| Stability metric | $E_{hull}$ |
| Stable-class criterion | $E_{hull} \leq 25$ meV $atom^{-1}$[21] |
| Classification target | Stability label: 1 = stable, 0 = unstable |
| Regression target | Scalar-relativistic PBE band gap, $E_g$ |
| Data split | 80:20 train:test, stratified by stability label[9] |
| Data content | $E_{hull}$, $E_g$, relaxed structures, composition metadata |

### 2.5 Chemical-genome descriptor engineering

Each $A_2BB'X_6$composition was encoded using a chemically interpretable descriptor set referred to as the chemical genome. The descriptor framework was designed not only to support predictive modeling, but also to preserve physical meaning so that the learned model behavior could be translated into chemical design rules.

The descriptor set was organized into six chemical-gene families. Geometric packing genes included ionic radii and tolerance-related descriptors (t, μ, τ). Framework cohesion genes comprised metal–halide bond-energy proxies and formation-energy-related descriptors. Chemical polarity/covalency genes included electronegativity differences and metal–halide polarity descriptors. Charge-transfer genes comprised ionization energy, electron affinity, and charge-accommodation descriptors. Polarization/screening genes included atomic polarizability and halide-softness descriptors. Electronic-identity genes comprised valence electron count, atomic number, and site-specific elemental identity descriptors[22–28].

Starting from the broader descriptor pool, redundant and weakly informative variables were removed using pairwise correlation analysis and feature-relevance screening[9]. Strongly collinear descriptors were pruned to avoid overrepresenting identical chemical information, while weakly relevant descriptors were excluded to maintain a compact and interpretable feature space. Descriptors were standardized to zero mean and unit variance before training when required by the learning algorithm. The final descriptor list and family assignment are provided in Table S3.

### 2.6 Machine-learning surrogate models and evolutionary optimization

Two supervised learning tasks were performed: thermodynamic stability classification and PBE band-gap regression. For stability classification, Decision Tree, Random Forest, and XGBoost classifiers were evaluated[29–32]. For band-gap prediction, Random Forest regression, support vector regression with a radial basis function kernel (SVR-RBF), and XGBoost regression were tested. Models were trained using the 1,221-compound DFT-labeled dataset and subsequently applied to the independently enumerated 13,088-composition Pb-free library. The DFT-labeled training dataset and the enumerated screening library were generated independently, although both belong to the halide double-perovskite family. The trained models thus transfer patterns learned from known DFT-labeled compounds to a broader chemically generated candidate space.

Optimizations for all hyperparameters were carried out using evolutionary strategy within a genetic algorithm. An initial population of random hyperparameter configurations was generated and improved using selection, crossover and mutation over 50 generations with a population of 50. The optimization goal for the stability classifier was maximizing recall for the stable class, since misses have greater costs than false positives in early stage materials discovery.

The band-gap regressor was optimized to minimize the mean squared error. For the thermodynamic stability model, we used a Decision Tree with these settings: max depth = 6, minimum samples to split = 5, minimum samples per leaf = 9, using the Gini index as the criterion, the best splitter, and max features set to 0.5699. For predicting the band gap, we used an XGBoost regressor with the following settings: learning rate = 0.0801, max depth = 8, 380 estimators, subsample = 0.9114, regularization parameters reg_lambda = 0 and reg_alpha = 0.3901, gamma = 0.3351, colsample_bylevel = 0.9534, and colsample_bytree = 0.9277. After fine-tuning, these models were retrained on the full training set and then applied to the entire compositional library for high-throughput screening (HTS), as shown in Table S2.

## 2.7 Model interpretation and backward mapping

Model interpretation was conducted to connect learned predictions back to chemically meaningful descriptor families. Feature-importance analysis and SHAP-based interpretation identified which chemical-gene families contributed most strongly to stability classification and band-gap regression[33–35]. The resulting trends were interpreted in terms of geometric packing, framework cohesion, polarity/covalency, charge transfer, polarization/screening, and electronic identity. This step is central to the backward-mapping strategy: the ML models were not treated as black-box predictors, but as sources of chemical insight identifying which variables control structural feasibility, stability tendency, and target band-gap behavior.

## 2.8 Staged screening and candidate prioritization

A staged inverse-design screening funnel reduced the 13,088-composition library to a DFT-validation set. In Stage I, the optimized stability classifier retained compositions predicted as stable. In Stage II, geometric formability filters based on $t$, $\mu$, and $\tau$ were applied. In Stage III, the optimized band-gap regressor selected candidates within the application-oriented target band-gap window. This procedure identified 113 ML-prioritized candidates within the stability–function region of interest, which were then subjected to DFT phenotype closure. DFT validation identified 10 phase-stable semiconductor candidates, from which 7 Pb-free double perovskites were retained for detailed discussion based on phase stability, band-gap relevance, chemical diversity, and descriptor consistency[36].

2.9 First-principles structural and electronic validation

First-principles calculations were performed in Materials Studio. Structural relaxations were carried out using $DMol^3$ within the spin-polarized GGA-PBE framework. A DNP 3.5 numerical atomic-orbital basis set was used together with scalar-relativistic DSPP pseudopotentials for heavy elements. Geometry optimization was continued until the total-energy change was below $1 \times 10^{-6}$Ha and the residual force was below $0.002$ Ha $Å^{-1}$. Both lattice parameters and internal atomic coordinates were allowed to relax. Monkhorst–Pack k-point meshes were selected based on convergence tests such that total energies, lattice constants, and band gaps were converged to within a few meV/atom, $0.01$Å, and $0.05$eV, respectively[37,38]. Spin polarization was included for all systems; for open-shell compounds containing Mn and Cu, calculations were initialized using high-spin ferromagnetic configurations. Because $DMol^3$ employs a numerical atomic-orbital basis rather than a plane-wave basis, calculated electronic properties were used primarily for internally consistent comparison and ranking within the same computational framework. Main conclusions are therefore based on relative trends across candidates rather than direct numerical comparison with values from other DFT codes.

Electronic band structures and density of states were calculated from the optimized geometries. The band gap, direct or indirect nature, and band-edge dispersion were extracted. The projected density of states (PDOS) helped us understand which electron orbitals are most important at the key energy levels—the valence-band maximum and conduction-band minimum. It gives us a clearer picture of how these orbitals influence the material's ability to conduct electricity.

2.10 Optical-property calculations and simulated XRD

We use CASTEP with GGA-PBE and norm-conserving pseudopotentials to calculate optical properties. We look at the complex dielectric function, $\varepsilon(\omega) = \varepsilon_1(\omega) + i\varepsilon_2(\omega)$. This help us get values like absorption coefficient, refractive index, extinction coefficient, reflectivity, energy-loss function, and optical conductivity. These calculations show us how the material absorb light, lose energy, reflect, plasmon-like behavior, and convert photons to charge[37–39].

Full optical spectra are provided in Figures S7–S12. Simulated powder XRD patterns were generated from the optimized structures to provide crystallographic fingerprints for future experimental comparison; these patterns serve as simulated structural fingerprints rather than experimental proof of phase formation. The optimized structures and simulated XRD patterns are provided in Figure S5.

2.11 Effective mass and transport analysis

Carrier effective masses were extracted from the curvature of the electronic bands near the conduction-band minimum and valence-band maximum. Parabolic fitting was applied near the band extrema, and electron and hole effective masses were obtained along relevant high-symmetry directions. Directional values were averaged to provide scalar effective-mass descriptors for comparative discussion.

For selected absorber-oriented candidates, phonon-limited mobilities were estimated using the deformation-potential approach. Band-edge shifts and elastic responses were extracted under small uniaxial strain, and mobilities were estimated using the 3D Bardeen–Shockley deformation-potential model:

$$\mu = \frac{2e\hbar^4 C}{3k_B T (m^*)^2 E_1^2}$$

where C is the elastic constant, $E_1$is the deformation potential, $m^*$is the carrier effective mass, Tis temperature, and the remaining symbols have their usual meanings. Direction-resolved transport data are provided in the Supporting Information, while direction-averaged values are used in the main discussion[40–44].

### 2.12 Chemical-genome design-rule extraction

We mapped DFT-validated structural, electronic, transport, optical, and XRD properties onto six chemical-gene families. Genes related to geometric packing were connected to the material's structural formability. Framework cohesion genes were tied to phase stability, while polarity/covalency genes influenced band-edge hybridization and optical-transition strength. Charge-transfer genes helped determine chemical plausibility and band alignment. Genes related to polarization and screening were linked to dielectric response, and finally, electronic-identity genes were associated with band dispersion and carrier effective mass. This mapping assigned each validated compound to its most appropriate device role and extracted transferable design rules for future Pb-free discovery[45–47].

## 3. Results and Discussion

### 3.1 Anatomy of the lead-free double-perovskite chemical space and halide-aware formability manifold

The backward-mapping workflow begins by defining the $A_2BB'X_6$composition space before any machine-learning prioritization is applied. This step is important because device-level target windows are only useful if they are interpreted within a chemically and structurally realistic search space. Therefore, the initial goal was not simply to generate a large number of formulas, but to construct a charge-balanced library that can plausibly support the ordered double-perovskite framework.

The library was created using two oxidation-state families; heterovalent $A_2B^+B^{3+}X_6$ and homovalent $A_2B^{2+}B^{2+}X_6$. For the A-site, the pool of elements included Li, Na, K, Rb, Cs, Cu, Ag, and Au. The B/B⁻-site pool consisted of cations from alkaline-earth metals, transition metals, post-transition metals, and metalloids.

The halide site was varied over F, Cl, Br, and I. After enforcing charge neutrality and removing duplicate homovalent $B/B'$permutations, this procedure produced 13,088 unique charge-balanced lead-free $A_2BB'X_6$compositions.

To evaluate whether these charge-balanced formulas are geometrically realistic, we calculated the Goldschmidt tolerance factor t, octahedral factor μ, and new tolerance factor τ, using Shannon ionic radii. The accepted formability windows were $0.80 \leq t \leq 1.10$, $0.414 \leq \mu \leq 0.732$, and $\tau < 4.18$[18–20]. These descriptors were used as the first structural screening gate, not as final thermodynamic-stability predictors. In the chemical-genome framework, they represent the packing genes, which determine whether a composition can enter the structurally admissible double-perovskite manifold before more expensive ML and DFT evaluation.

Table 2. Enumeration rules and geometric formability criteria used to construct the $A_2BB'X_6$ library.

| Category | Definition used in this work |
| --- | --- |
| Structural formula | $A_2BB'X_6$ |
| Oxidation-state families | $A_2B^+B'^{3+}X_6$; $A_2B^{2+}B'^{2+}X_6$ |
| A-site pool | Li, Na, K, Rb, Cs, Cu, Ag, Au |
| $B/B'$-site pool | Be–Ba, Mn–Zn, Pd, Ir, Pt, Al, Ga, In, Sn, Ge, Sb, Bi, Te, Pb |
| Halide pool | F, Cl, Br, I |
| Excluded element | Pb |
| Duplicate handling | Homovalent $B/B'$permutations canonicalized to remove duplicate formulas |
| Geometric formability gates | $0.80 \leq t \leq 1.10$; $0.414 \leq \mu \leq 0.732$; $\tau < 4.18$ |
| Final enumerated library | 13,088 unique charge-balanced lead-free compositions |

The resulting library is not uniformly distributed across B–B′ chemistry. As shown in Figure 2a, the combinatorial frequency map reveals denser regions for several cation-pair families, reflecting the combined effects of oxidation-state availability, charge balance, and compatibility with multiple A-site and halide choices. This demonstrates that the searchable chemical space is intrinsically structured before ML screening begins: some regions are chemically accessible by enumeration, while others are inherently sparse.

The t–μ formability projection (Figure 2b) further shows that charge balance alone is insufficient to define a plausible double-perovskite candidate. Many nominally valid compositions fall outside the feasible packing manifold once geometric constraints are applied. The retained region corresponds to compositions where A-site cage filling and B-site octahedral compatibility are simultaneously satisfied, justifying packing descriptors as a hard pre-filter: candidates outside this region may appear electronically attractive but are unlikely to remain structurally meaningful after relaxation.

Formability statistics show a clear dependence on the type of halide (Figure 2c). Chloride- and bromide-based compositions have the highest retained fractions, 48.2% and 45.8%, respectively. Fluoride and iodide compositions are much less.  It retained only 12.5% and 15.4% respectively. This makes sense chemically: fluoride frameworks are often too compact, which over-constrains the octahedral network. Iodide frameworks expand the lattice too much, pushing many cation combinations out of the optimal tolerance range. Chloride and bromide offer a better balance, allowing a wider variety of double-perovskite compositions.

This analysis shows that the lead-free double-perovskite search space is not only large but also heavily influenced by oxidation-state chemistry, cation-size matching, and halide-dependent packing. The initial 13,088 charge-balanced formulas are just the starting point. The physically meaningful region is a smaller subset that meets the formability criteria.

This yields the first design rule of the backward-mapping workflow: structural feasibility must be enforced before functional optimization. Subsequent ML screening and DFT validation are therefore conducted within a chemically disciplined space where packing-controlled formability constitutes the first interpretable layer of the chemical genome.

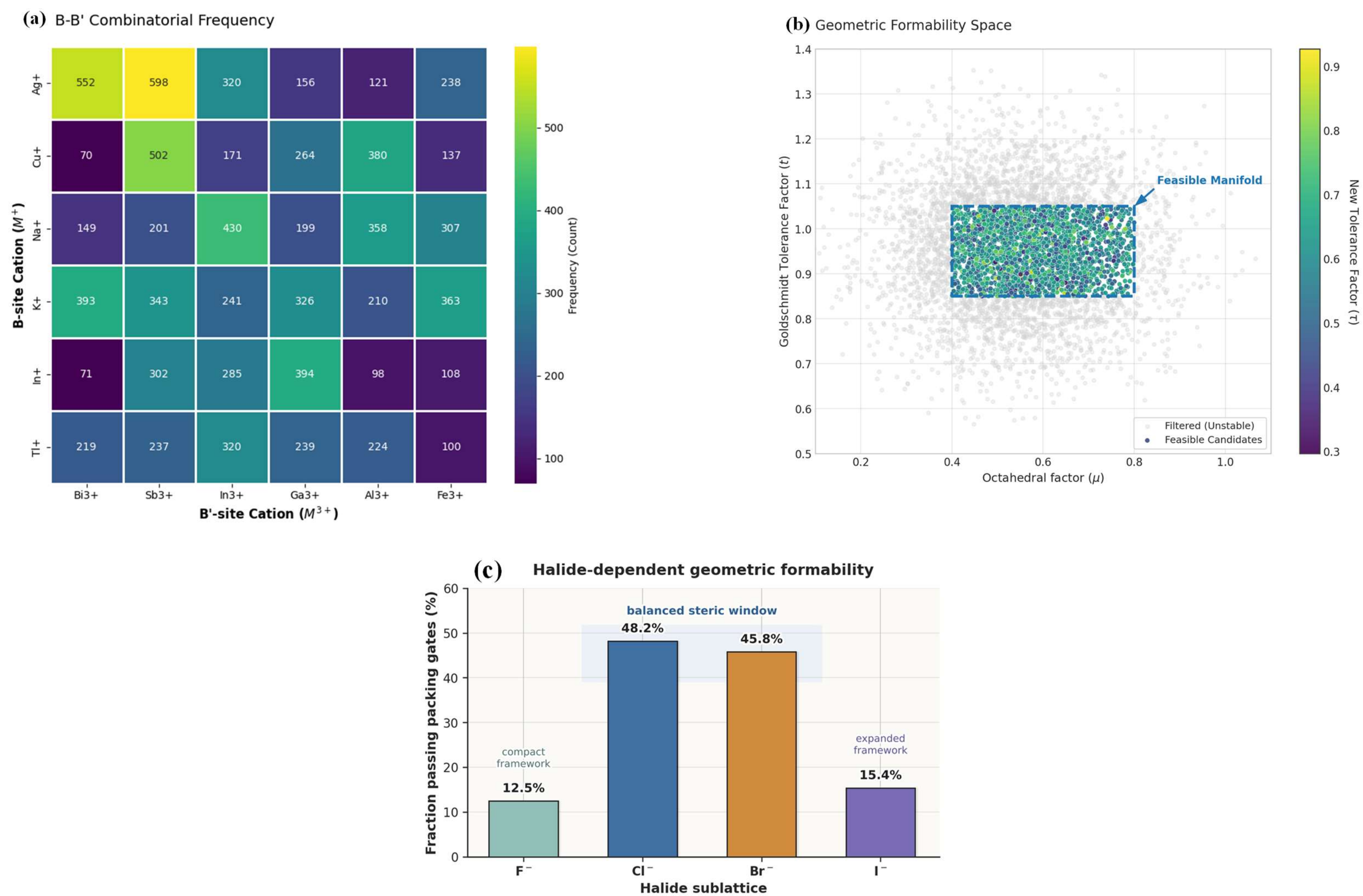

Figure 2. Chemical-space anatomy and halide-aware formability manifold of the enumerated lead-free library. (a) B–B′ combinatorial frequency map aggregated overall A-site and halide choices. (b) Geometric formability projection in t–μ space, showing the feasible packing manifold defined by $0.8 \leq t \leq 1.1$, $\mu \geq 0.41$, and $\tau < 5.0$. (c) Halide-dependent fraction of compositions passing geometric formability filters.

### 3.2 Chemical-genome descriptor construction and feature pruning

To make the backward-mapping workflow interpretable, the initial descriptor pool was organized into six physically motivated chemical-gene families rather than treated as an unstructured feature list. This classification separates the principal chemical factors controlling structural formability, thermodynamic stability, band-gap formation, dielectric response, and carrier transport.

Geometric packing genes include ionic radii and tolerance-factor descriptors such as t, μ, and τ, which describe A-site cage filling and B/B′-site octahedral compatibility. Framework cohesion genes, including A–X, B1–X, and B3–X bond dissociation energies and atomic formation enthalpies, describe metal–halide bond strength and lattice stability. Chemical polarity and covalency genes, primarily electronegativity-difference descriptors, encode bond polarity, covalency, and metal–halide hybridization. Charge-transfer genes, such as ionization energy and electron affinity, represent cation oxidation tendency and anion charge accommodation. Polarization and screening genes, based on how each site can be polarized, capture the material's lattice softness and dielectric response. Electronic-identity genes, which include the valence electron count and atomic number, help explain orbital filling, periodic trends, and the character of the band edges.

Table 3. Chemical-genome descriptor families.

| Chemical-gene family | Representative descriptors | Site or interaction | Physical interpretation | Role in the workflow |
|---|---|---|---|---|
| Geometric packing genes | Ionic radii R-A, R-B1, R-B3, R-X; Tolerance factors t, μ, τ | A, B, B′, X | Ionic size matching, A-site cage filling, octahedral compatibility, lattice distortion tendency | Defines the first structural formability gate |
| Framework cohesion genes | Bond dissociation energies A-X, B1-X, B3-X; Atom Formation enthalpy E-A, E-B1, E-B3, E-X | A, B, B′, X | Metal–halide bond strength, lattice cohesion, elemental stability tendency | Supports thermodynamic-stability learning |
| Chemical polarity and covalency genes | Electronegativity X-B1, X-B3, X-X | B, B′, X | Bond polarity, covalency, metal–halide hybridization, bonding asymmetry | Helps tune band-edge separation and optical-transition strength |
| Charge-transfer genes | First Ionization energy I-A; Atom Formation enthalpy | A, B, B′, X | Cation ionization tendency, anion | Connects elemental chemistry to |
|  | E-A, E-B1, E-B3, E-X |  | reducibility, charge redistribution | stability tendency and band alignment |
| Polarization and screening genes | Electronegativity X-B1, X-B3, X-X | A, B, B′, X | Lattice softness, dielectric screening, refractive response | Informs optical screening and exciton-shielding potential |
| Electronic-identity genes | Valence electrons V-A, V-B1, V-B3, V-X; Atomic number N-A, N-B1, N-B3 | A, B, B′ | Valence filling, orbital identity, periodic trend, relativistic character | Controls band-edge chemistry, effective mass, and transport asymmetry |

Starting from the broader descriptor pool, feature pruning was performed to reduce redundancy and improve interpretability. Pairwise correlation analysis was used to remove strongly collinear variables, while descriptors with very weak correlation to both target phenotypes, $| r | < 0.10$with respect to $E_g$and the $E_{hull}$-derived stability label, were excluded. This threshold served as a practical filter rather than a causal criterion, maintaining a compact descriptor set while preserving chemically meaningful information.

As shown in Figure 3a, $E_g$ and the $E_{hull}$-derived stability label are controlled by different descriptor subsets, confirming that band-gap formation and thermodynamic stability are related but not identical learning targets. Bonding descriptors, especially B1–X and B3–X bond-energy terms, show strong relevance to band-gap variation, consistent with the role of metal–halide hybridization in defining the band edges. The retained descriptor ranking (Figure 3b) further shows that important variables span multiple gene families, indicating that the model does not rely on a single geometric or electronic proxy.

The pruning step thus converts the initial descriptor pool into a compact chemical genome that is both predictive and interpretable. Packing genes define the structurally admissible manifold; cohesion and polarity genes describe the bonding environment; charge-transfer and polarization genes encode electronic response; and electronic-identity genes connect composition to band-edge character and carrier transport. This descriptor architecture provides the genotype layer for subsequent ML screening and enables chemically traceable interpretation of predicted stability and optoelectronic trends.

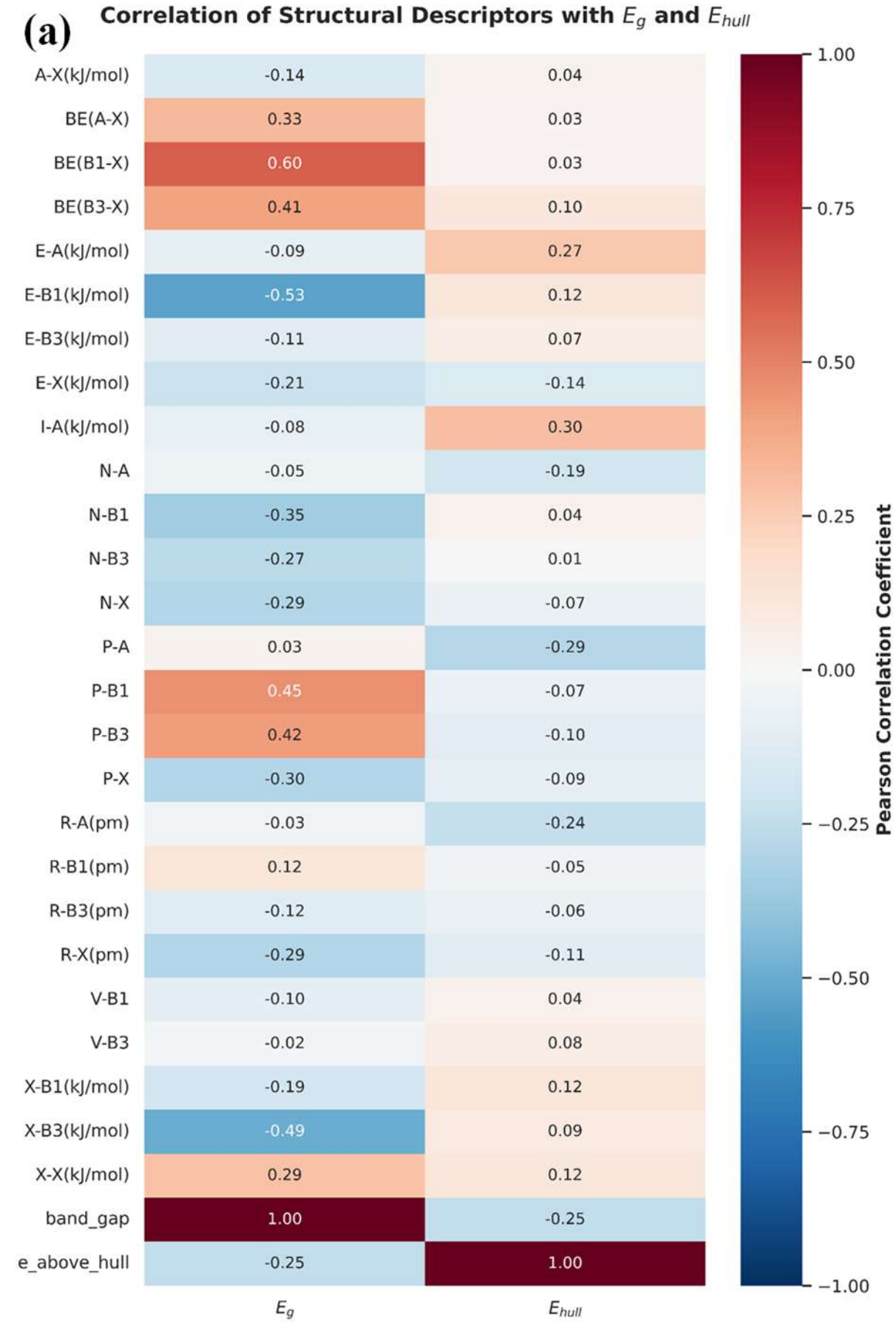

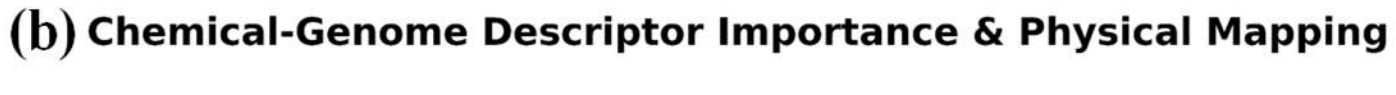

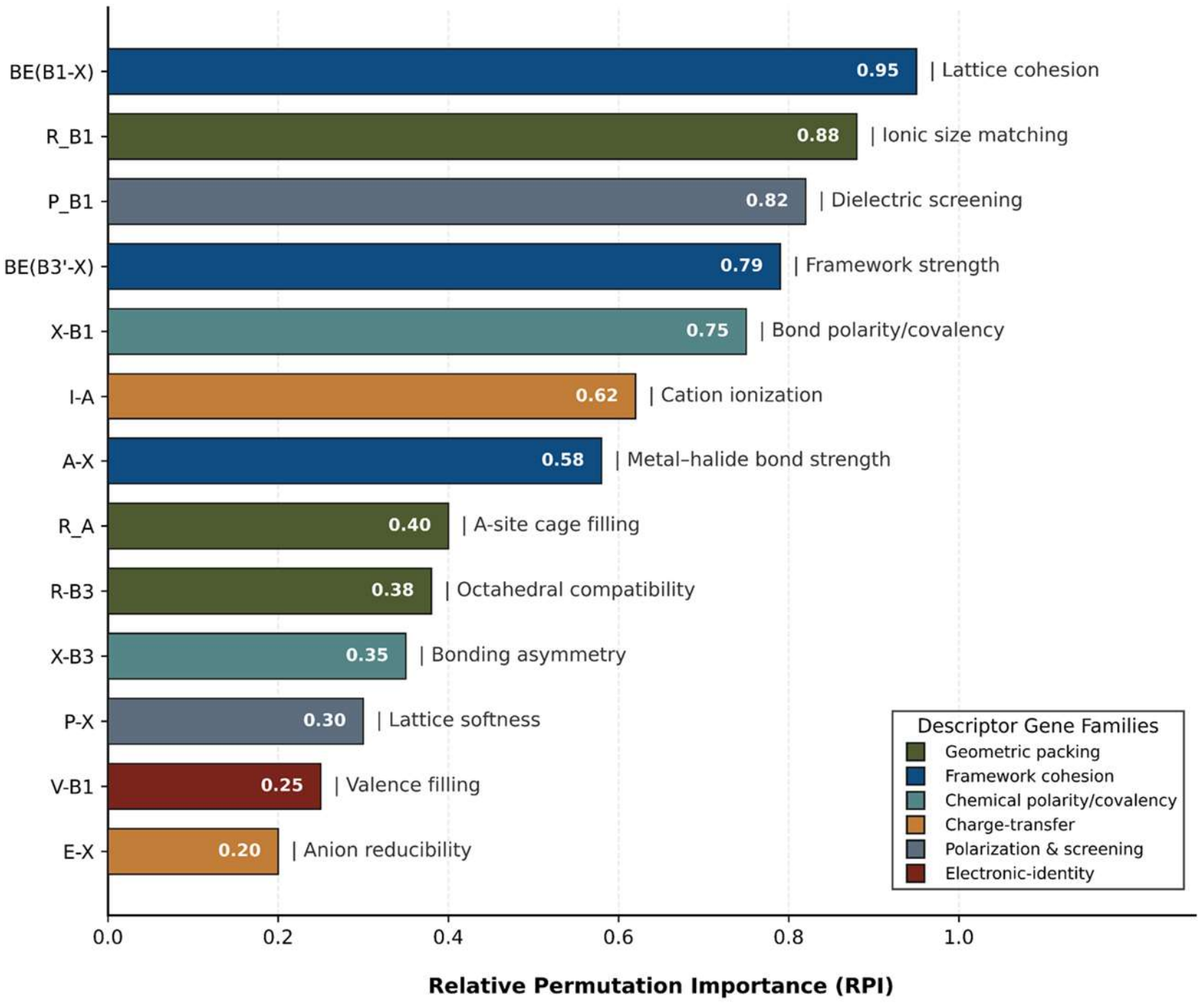

Figure 3.Chemical-genome descriptor pruning and physical interpretability. (a) Descriptor–target correlation matrix for $E_g$ and the $E_{hull}$-derived stability label. (b) Retained descriptor set ranked

by relative permutation importance and colored by chemical-gene class, showing that stability and band-gap prediction require coupled steric, bonding, and electronic information.

### 3.3 Genome-to-phenotype machine-learning surrogate performance

After constructing the chemical-genome descriptor space, we evaluated whether the retained descriptors can reliably translate composition-level chemistry into DFT-derived phenotypes. In this work, the ML models serve as genome-to-phenotype surrogate maps that rapidly prioritize the 13,088-member lead-free chemical library before DFT validation, not as final replacements for first-principles calculations. The models were therefore assessed not only by conventional accuracy metrics, but also by their usefulness in a screening workflow where retaining promising candidates is more critical than prematurely rejecting them.

Two supervised learning tasks were considered: thermodynamic-stability classification ($\mathrm{E_{hull}} \leq 25\mathrm{meV\ atom^{-1}}$ labeled as stable) and band-gap regression (targeting scalar GGA-PBE band gap $\mathrm{E_g}$). These phenotypes were modeled separately because thermodynamic accessibility and electronic-gap formation are related but distinct materials properties.

For thermodynamic stability classification, we tested evolutionary-algorithm (EA)-optimized Decision Tree, Random Forest, and XGBoost classifiers (see Table 4). The EA-optimized XGBoost classifier showed the highest overall accuracy (0.860) and stable-class precision (0.865). The EA-optimized Decision Tree was selected as the primary stability-screening model. As, it produced the highest stable-class recall (0.831). This choice is appropriate for inverse screening because a false positive only increases the cost of downstream DFT validation, whereas a false negative permanently removes a potentially stable compound from the discovery funnel.

Table 4. Test-set performance of the evolutionary-optimized stability classifiers.

| Model | Accuracy | Precision (stable) | Recall (stable) | Screening interpretation |
|---|---|---|---|---|
| Random Forest, EA-optimized | 0.846 | 0.835 | 0.792 | Balanced baseline |
| XGBoost, EA-optimized | 0.860 | 0.865 | 0.792 | Highest overall accuracy |
| Decision Tree, EA-optimized | 0.833 | 0.785 | 0.831 | Selected for recall-prioritized screening |

The confusion matrix confirms that the selected Decision Tree preserves a large fraction of stable candidates while rejecting many unstable compositions. Precision–recall and Receiver Operating Characteristic (ROC) curve indicate strong separability between stable and unstable classes, while calibration behavior shows that predicted stability probabilities are most meaningful in the intermediate-to-high probability region. This is sufficient because the classifier functions as a realization-stage filter before DFT phenotype closure, not as a final thermodynamic authority.

For band-gap regression, the EA-optimized XGBoost regressor delivered the strongest overall performance ($R^2 = 0.9317$, $RMSE = 0.5144eV$, and $MSE = 0.2646eV^2$) (see Table 5). The parity plot shows predicted and DFT−calculated $E_g$ values remain close to the identity line across the 0–8 eV range. The residual distribution is centered near zero with limited systematic bias, and the error-versus-DFT-$E_g$ plot demonstrates reasonable stability across both low-gap and wide-gap regimes. The evolutionary convergence curve confirms the final model reached a stable low-error region during hyperparameter optimization.

Table 5. Test-set performance of the evolutionary-optimized band-gap regressors.

| Model | $R^2$ | RMSE (eV) | MSE ($eV^2$) | Screening interpretation |
| --- | --- | --- | --- | --- |
| SVR, RBF, EA-tuned | 0.9244 | 0.5414 | 0.2931 | Strong nonlinear baseline |
| Random Forest, EA-tuned | 0.9241 | 0.5426 | 0.2944 | Ensemble baseline |
| XGBoost, EA-tuned | 0.9317 | 0.5144 | 0.2646 | Selected band-gap surrogate |

The XGBoost regressor should be interpreted as a screening-level band-gap surrogate, not as a substitute for final DFT electronic-structure analysis. Its role is to identify compositions whose predicted gaps fall within target-relevant windows, thereby reducing the number of candidates requiring first-principles evaluation. This distinction is important because PBE-level band gaps are approximate, and final candidates still require DFT validation of band structure, orbital character, carrier effective mass, and optical response.

Interpretability analysis indicates that the surrogate models learned chemically meaningful relationships. Permutation importance and partial-dependence analysis identify BE(B1–X), $P_{B1}$, X–B1, $R_{B1}$, and BE(B3–X) among the most influential descriptors controlling predicted band gap (see Figure 6). These variables correspond to framework cohesion, local polarizability, bond polarity, octahedral fit, and metal–halide hybridization, consistent with the chemical-genome framework where bonding, polarization, and electronic-identity descriptors jointly regulate band-edge formation. Two-feature interaction maps further show that $E_g$ is not controlled by a single descriptor independently, but by coupled changes in metal–halide bonding and B-site polarizability.

Overall, the surrogate-learning results show that the selected chemical-genome descriptors are useful for large-scale screening. They also remain physically interpretable.

The Decision Tree classifier protects the discovery funnel from premature loss of potentially stable candidates by prioritizing recall, whereas the XGBoost regressor provides accurate screening-level band-gap prediction. Together, these models form the quantitative bridge between descriptor-space backward mapping and composition-space realization.

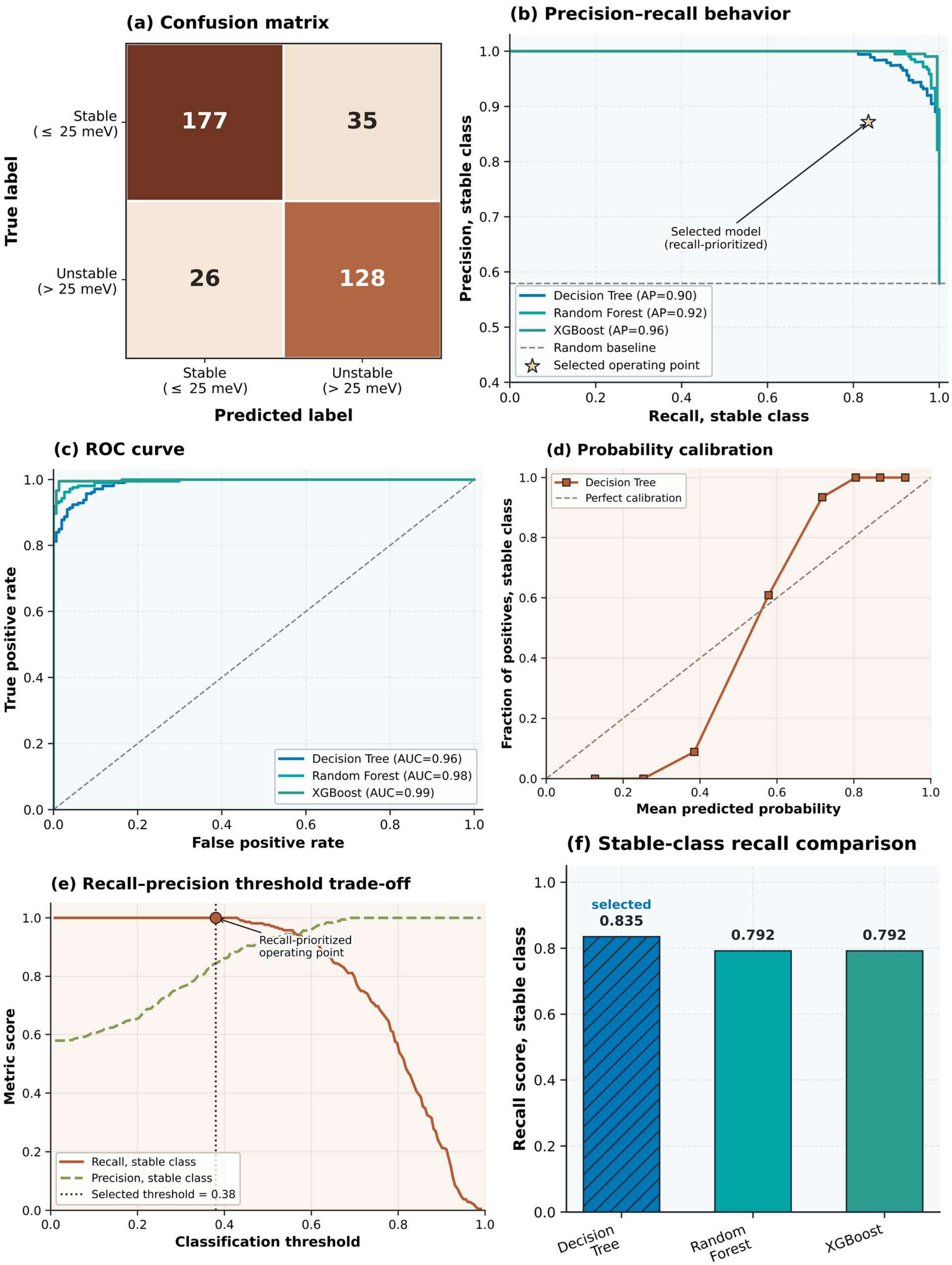

Figure 4. Shows the performance of the recall-prioritized stability classifier. It includes the confusion matrix, precision-recall curve, and Receiver-operating characteristic curve (ROC) curve. The figure also displays probability calibration and operating-threshold analysis. Finally, it compares the recall for the selected EA-optimized Decision Tree classifier ($E_{hull} \leq 25$meV $atom^{-1}$).

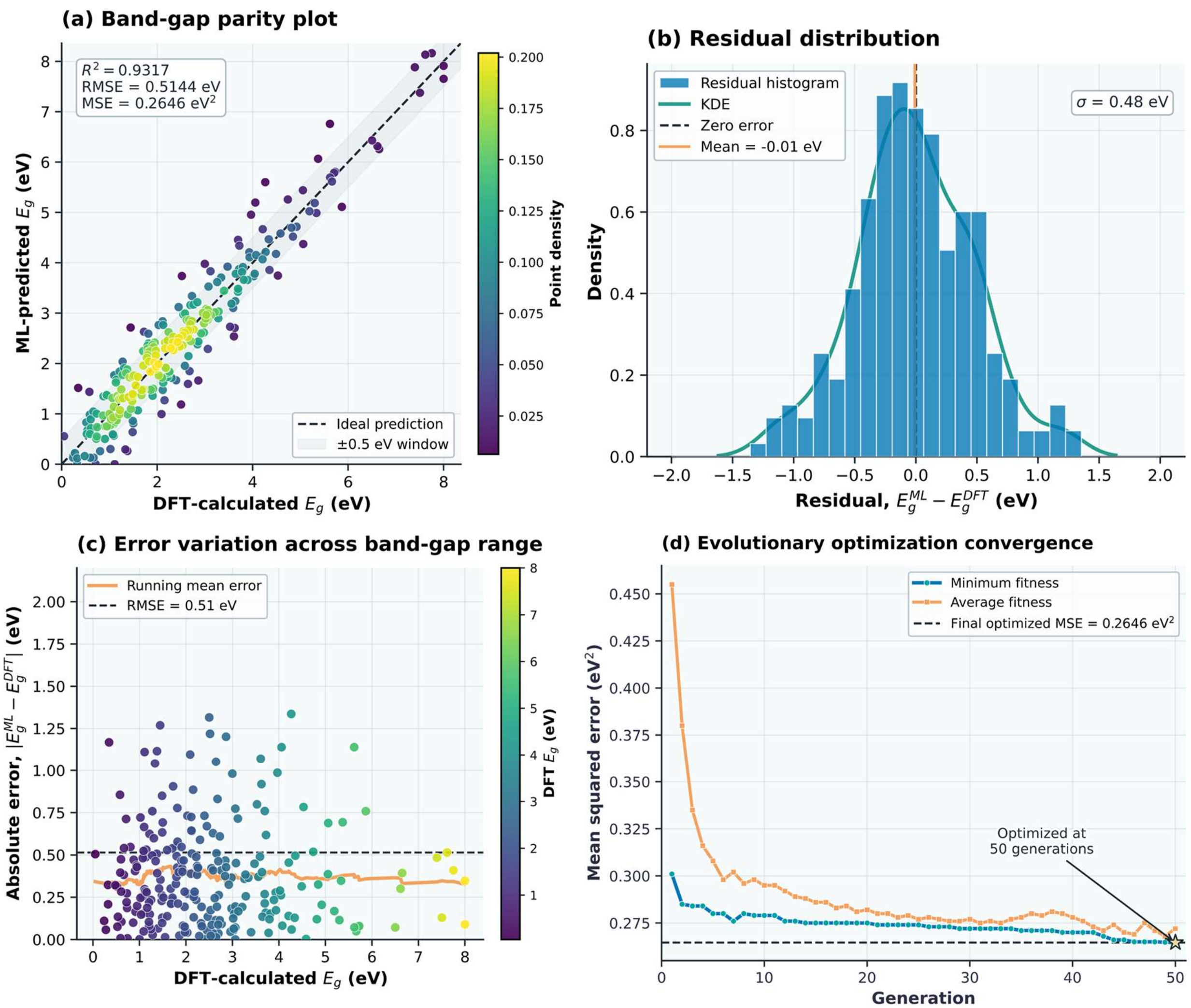

Figure 5. Band-gap surrogate model performance of the EA-optimized XGBoost regressor. It includes the parity plot, residual distribution, and error-versus-DFT-$E_g$ panel. The figure also displays validation MSE convergence during evolutionary optimization. The performance metrics are $R^2 = 0.9317$, RMSE = 0.5144 eV, and MSE = 0.2646 eV$^2$.

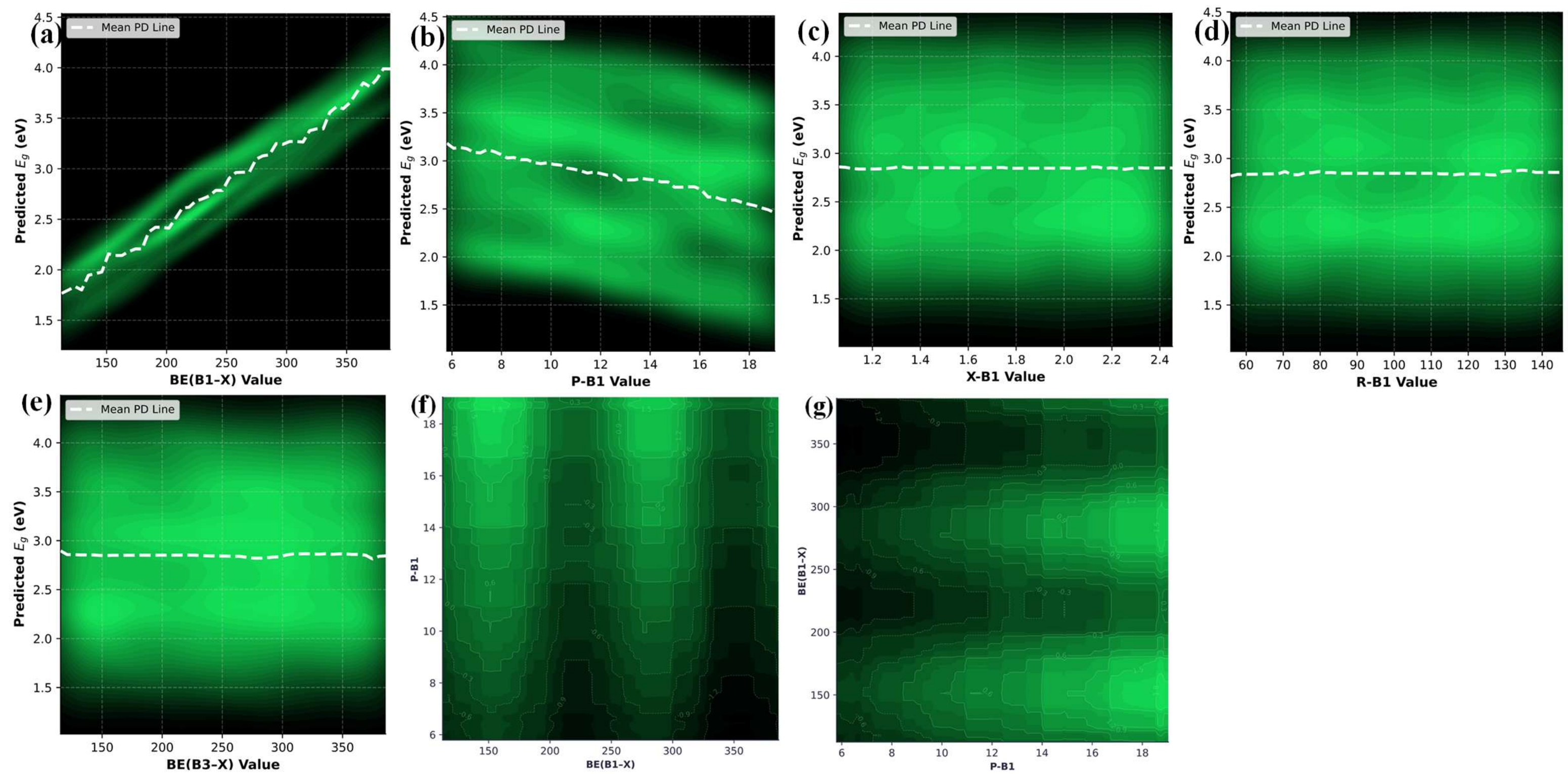

Figure 6. Partial dependence analysis of the top five descriptors controlling the predicted band gap. The one-dimensional PDPs show how each feature affects the band gap, ordered by importance: (a) BE(B1–X) (rank 1), (b) P–B1 (rank 2), (c) X–B1 (rank 3), (d) R–B1 (rank 4), and (e) BE(B′–X) (rank 5). (f) Shows a two-dimensional interaction PDP, highlighting how the two highest-ranked descriptors, BE(B1–X) and P–B1, influence the predicted band gap. (g) Maps the DFT-calculated band gaps over the P–B1–BE(B1–X) space, validating the model's trends against first-principles results. Color bars show the band-gap values (in eV), and contours in (f) and (g) help visualize the response gradients and interaction effects.

### 3.4 Mechanistic genome decoding for backward target mapping

After validating the genome-to-phenotype surrogate models, we examined whether the learned model behavior could be translated into chemically meaningful design rules—the core of the backward-mapping strategy. A useful inverse-design model should not only predict which compositions may satisfy a target window but also reveal which chemical variables move a composition toward structural stability and useful optoelectronic behavior.

Figure 7 summarizes this mechanistic decoding. The SHAP distributions (see Figure 7a, b) highlight the descriptors that have the strongest influence on predicted stability and band-gap phenotypes. The response curves (Figure 7c–e) then translate these effects into physical design rules. Together, these analyses uncover a layered genome hierarchy: packing defines formability first, bonding controls framework cohesion next, and electronic-response descriptors ultimately tune the final band-gap behavior. For the stability classifier, the SHAP distribution (Figure 7a) shows that the stable/unstable decision is not governed by a single descriptor but depends on coupled A-site response, formation-enthalpy, ionic-size, and halide-chemistry descriptors, as listed in Table 6. Stability emerges from the combined action of packing, charge-transfer, halide-response, and framework-cohesion genes.

Table 6. SHAP-decoded chemical-genome rules for backward target mapping.

| Target or genome layer | Dominant SHAP descriptors | Mechanistic interpretation | Backward-mapping rule |
|---|---|---|---|
| Stability classifier | I-A, P-A, E-A, R-A, E-X, X-B1, X-X, R-X | Stability is controlled by coupled A-site response, halide chemistry, size compatibility, and formation-energy terms. | Enforce A-site/halide compatibility and favorable framework chemistry before device-function filtering. |
| Band-gap regressor | B1-X, E-B1, X-B3, P-B1, P-B3, B3-X, N-B1, P-X, R-X | $E_g$ is influenced mainly by B/B′–X bonding, polarizability, electronegativity, and halide-mediated lattice response. | Tune $E_g$ through octahedral-site chemistry rather than global composition changes alone. |
| Packing gate | τ, R-A, R-X | Stability probability decreases outside the favorable formability window. | Treat packing feasibility as the first hard constraint. |
| Cohesion control | B-X, B1-X, B3-X, E-B1, E-X | Stronger metal–halide bonding supports structural persistence. | Optimize cohesion only inside the formability manifold. |

| Opto-electronic tuning | X-B, P-B1, P-B3, N-B1, N-B3, R-X | Band-edge separation and screening depend on electronegativity, polarizability, and orbital identity. | Use B/B′-site identity and halide response as target-specific $E_g$ knobs. |
|---|---|---|---|

The band-gap regressor exhibits a different descriptor hierarchy (see Figure 7b). Dominant descriptors are mainly associated with the octahedral sublattice, including B–X and B′–X bond energies, B-site polarizability, B-site electronegativity, and B-site ionic radius. This indicates that $E_g$ is controlled most directly by the strength, polarity, and polarizability of the metal–halide framework. Structural feasibility is required for realization, but band-gap tuning occurs primarily through octahedral bonding and electronic-response genes. The response curves (see Figure 7c–e) make this hierarchy explicit. The predicted stability probability decreases sharply when the new tolerance factor τ moves beyond the favorable formability region, confirming packing as a feasibility gate. Once this geometric condition is satisfied, B-site metal–halide bond dissociation energy becomes a cohesion-control variable: stronger B–X bonding generally supports higher stability probability. Finally, predicted band gap increases with B-site electronegativity, showing that octahedral cation identity can tune band-edge separation. However, this tuning must be interpreted together with bond energy, polarizability, and ionic-radius effects rather than as an isolated descriptor trend.

These results convert model interpretability into a practical backward-mapping sequence: compositions must remain inside the formability manifold; the metal–halide framework must provide sufficient cohesive strength; and the band gap should be tuned through octahedral-site descriptors such as B-site electronegativity, polarizability, bond strength, and ionic radius. This explains why single-property ranking is insufficient: a candidate with an attractive predicted band gap may still fail if it lies outside the packing or bonding window. The SHAP and response analyses show that the chemical genome behaves as a set of linked control layers connecting composition to structural feasibility, stability, and optoelectronic function—the principal value of the backward-mapping framework.

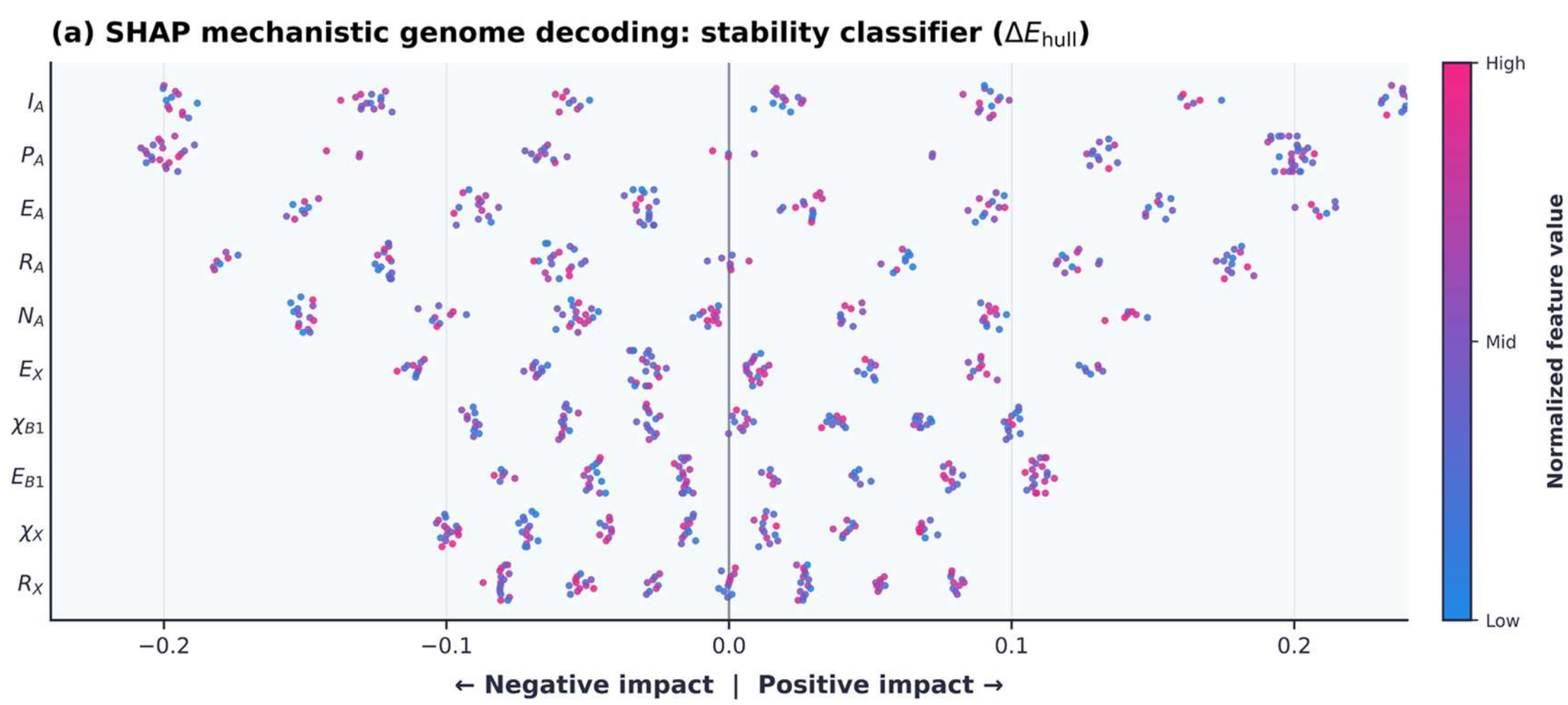

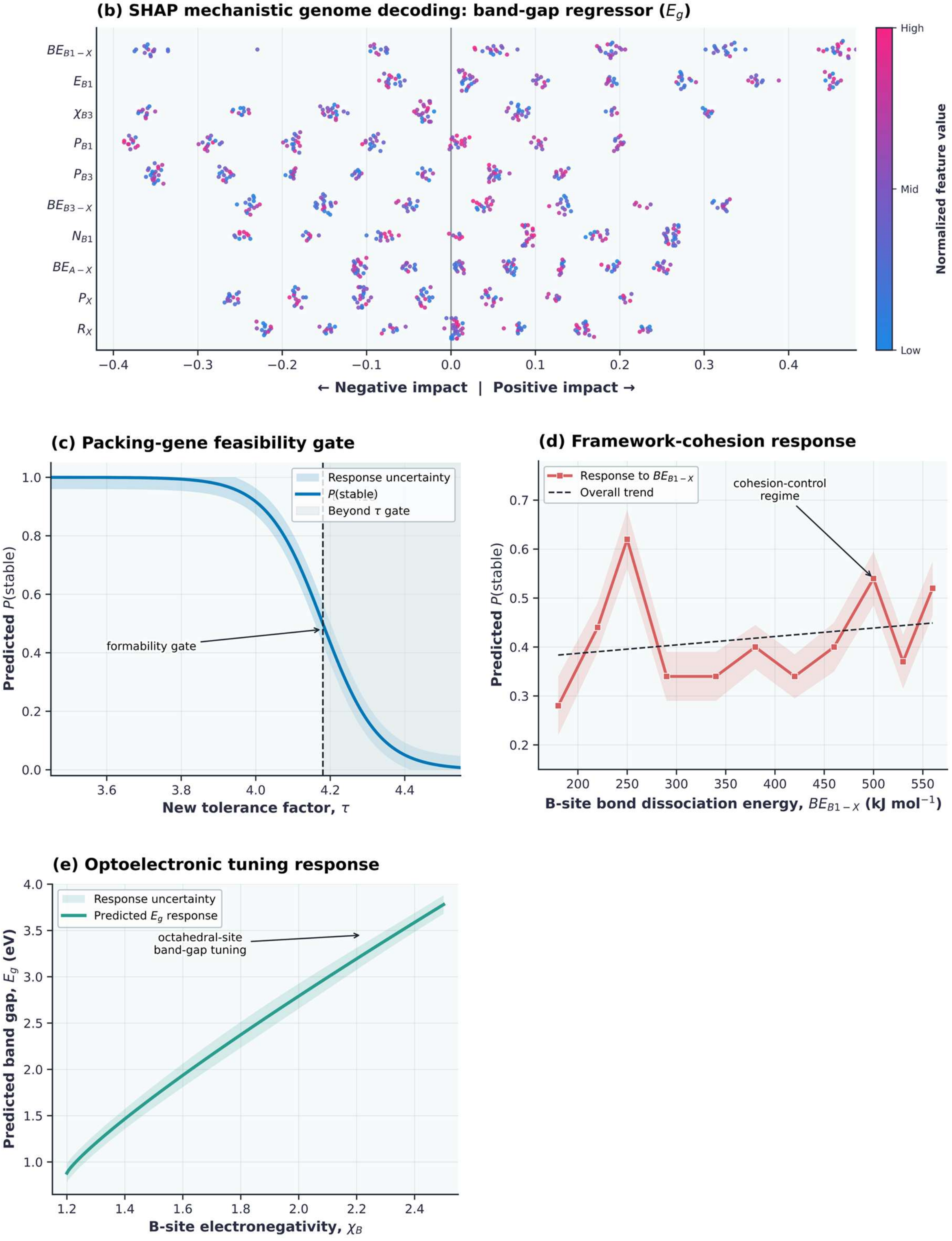

Figure 7. Mechanistic genome decoding for backward target mapping. (a) SHAP summary for stability classifier. (b) SHAP summary for band-gap regressor. (c) Stability probability vs. τ. (d) Stability response to B-site metal–halide bond dissociation energy. (e) Band-gap response to B-site electronegativity.

### 3.5 Inverse-design screening funnel and stability–function intersection

After constructing the chemical-genome descriptors and validating the surrogate models, the workflow was applied to the full lead-free library to translate descriptor-level rules into candidate compositions. Screening was treated as a staged inverse-design funnel where each stage imposes a different physical or functional requirement, so the final candidates represent the overlap between charge balance, predicted stability, structural formability, target band-gap placement, and DFT-validated semiconductor behavior (see Table 7 and Figure 8).

Table 7. Multi-stage inverse-design screening criteria and candidate reduction statistics.

| Screening stage | Candidates remaining | Fraction of initial (%) | Main criterion | Physical role in screening |
|---|---|---|---|---|
| Initial library | 13,088 | 100.00 | Charge-balanced $A_2BB'X_6$ enumeration | Defines the search space |
| Stage I | 5,354 | 40.91 | ML stability classifier | Preserves predicted stable compositions |
| Stage II | 733 | 5.60 | Geometry screening using $t$, $\mu$, and $\tau$ | Enforces structural formability |
| Stage III | 113 | 0.86 | Application-oriented $E_g$ targeting | Identifies the ML-prioritized stability–function region |
| Stage IV | 10 | 0.08 | DFT phenotype closure | Validates phase-stable semiconductor candidates |
| Stage V | 7 | 0.05 | Final Pb-free discussion set | Retains candidates for detailed analysis |

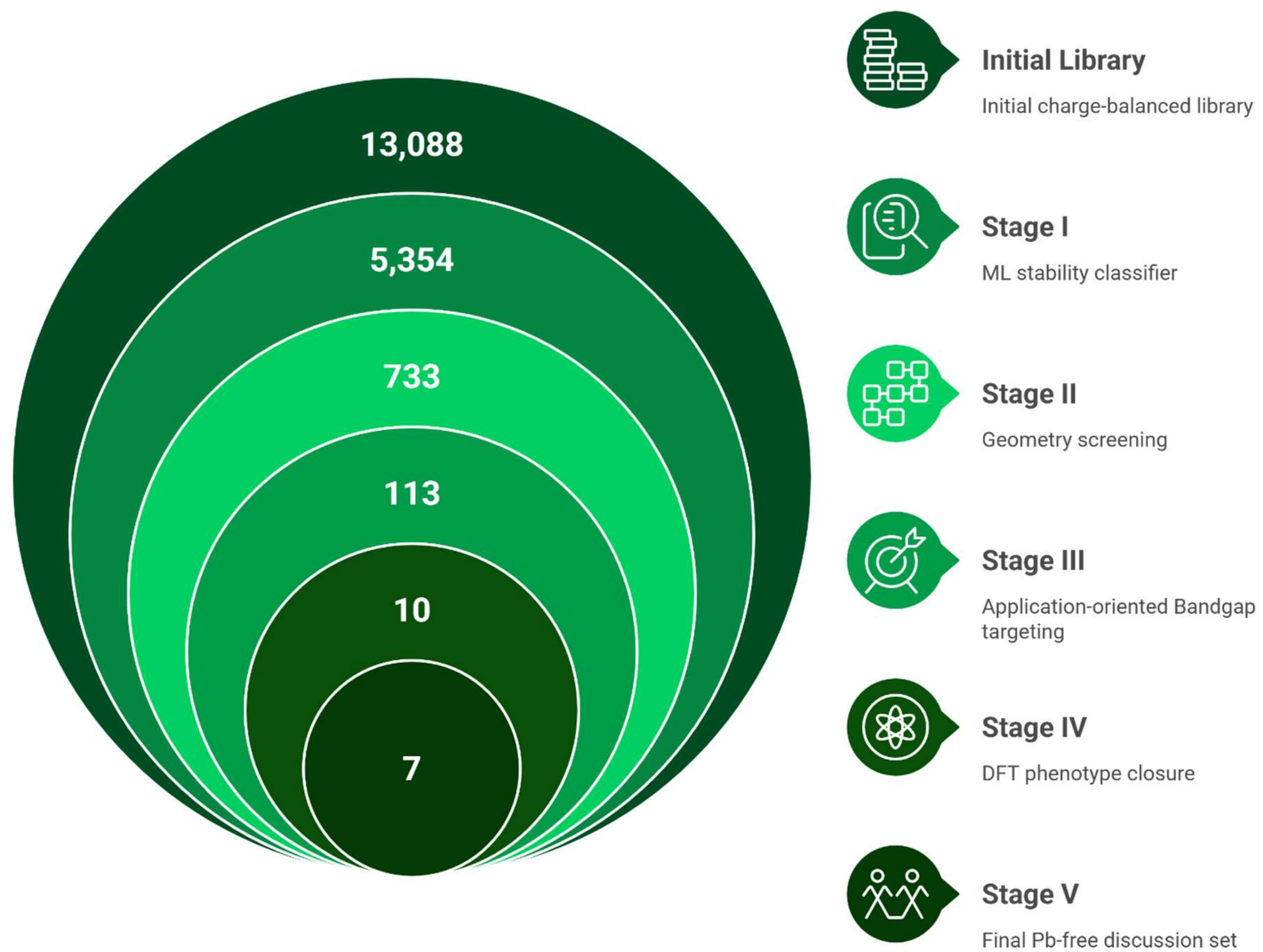

Figure 8. Inverse-design screening funnel. Sequential reduction of the 13,088-member library through ML stability prediction, geometric formability screening, band-gap targeting, DFT phenotype closure, and final Pb-free candidate selection.

The screening began with 13,088 charge-balanced $A_2BB'X_6$compositions. The recall-prioritized ML stability classifier retained 5,354 candidates (40.91%), preserving the broad predicted stable region (see Table7 and Figure 8). Geometry filtering using t, μ, and τ reduced the pool to 733 candidates (5.60%), demonstrating that structural formability is a major bottleneck in the lead-free double-perovskite space. The application-oriented band-gap filter further reduced the set to 113 candidates (0.86%), defining the ML-prioritized stability–function region. These candidates were selected as chemically enriched targets for first-principles validation, not as final stable compounds. DFT phenotype closure identified 10 phase-stable semiconductor candidates (0.08%), from which 7 final Pb-free double perovskites (0.05%) were retained for detailed discussion based on chemical diversity, band-gap relevance, and descriptor consistency. The reduction from 13,088 compositions to 7 final candidates highlights the narrowness of the stability–function intersection in $A_2BB'X_6$chemistry (see Table7 and Figure 8).

The stability–function landscape (see Figure 9) explains why this staged funnel is necessary. Many compositions may satisfy charge balance or predicted stability, but far fewer simultaneously fall within the target band-gap range and remain viable after DFT relaxation. Conversely, some compositions may show attractive predicted $E_g$ values but fail stability or formability requirements.

The final candidates thus occupy a small region where geometric feasibility, predicted thermodynamic accessibility, and target-relevant electronic response are satisfied simultaneously. The ML models are best understood as prioritization tools defining a chemically enriched region for DFT testing, while DFT provides final phenotype closure.

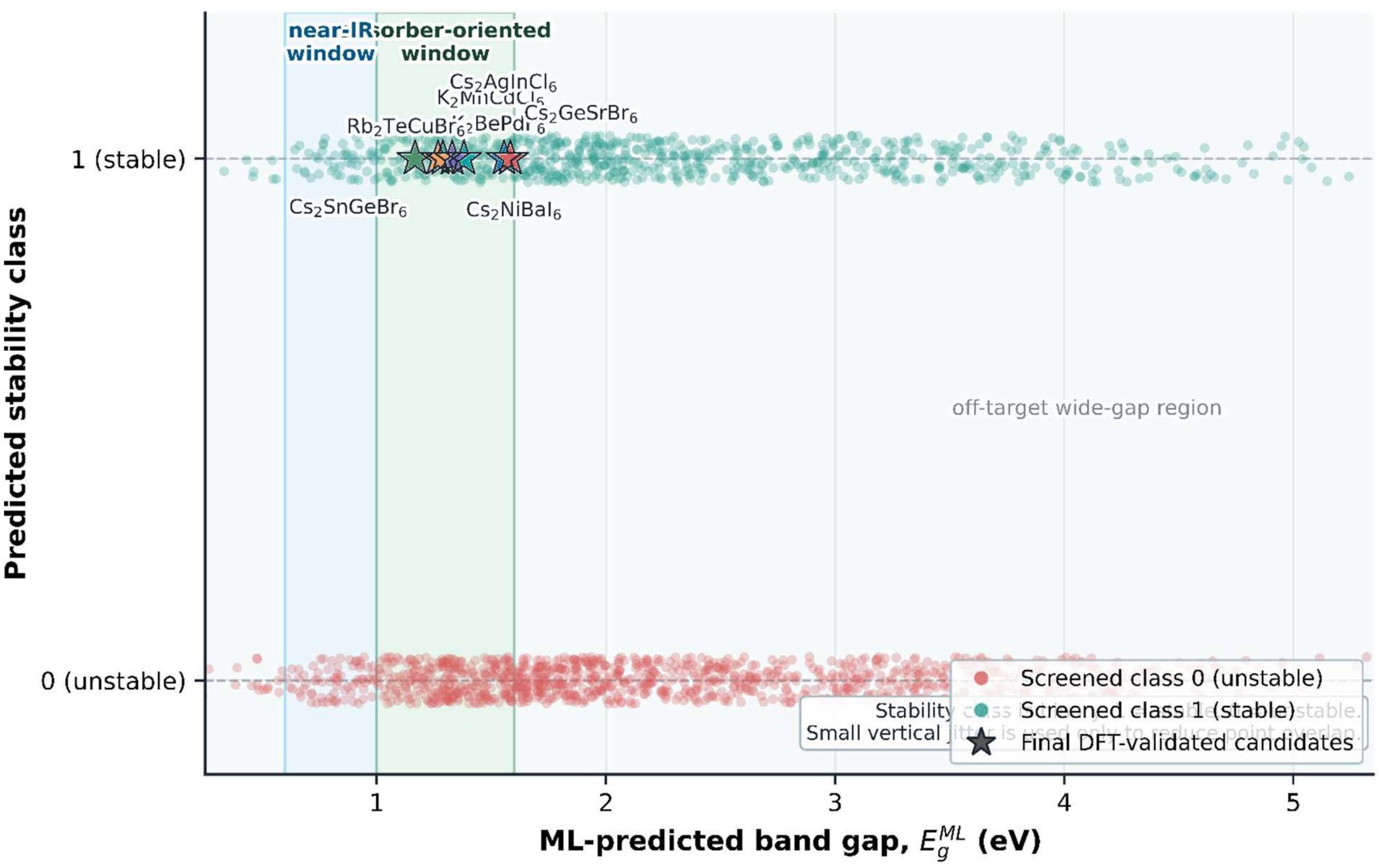

Figure 9. Stability–function landscape. Predicted stability probability vs. ML-predicted band gap with target application windows and highlighted candidates selected for DFT validation.

## 3.6 DFT-validated phase-stable semiconductor candidates

First-principles validation closes the loop between screening predictions and physically meaningful materials phenotypes. At this stage, the question is whether the shortlisted compounds remain ordered, near-hull, semiconducting, and optically credible after full structural and electronic relaxation.

In the final lead-free scope of this work, the retained compounds are $K_2BePdF_6$, $K_2MnCdCl_6$, $Rb_2FeMnF_6$, $Rb_2TeCuBr_6$, $Cs_2SnGeBr_6$, $Cs_2GeSrBr_6$, $Cs_2NiBaI_6$, and $Cs_2AgInCl_6$ (see Table 8). These represent the Pb-free validated endpoint of a much broader screened $A_2BB'X_6$ chemical space. Although broader workflow exploration also yielded chemically informative comparator systems, Pb-containing compounds were excluded from the final discovery set to maintain consistency with the lead-free scope. The validated Pb-free set spans different semiconducting phenotypes: $Cs_2SnGeBr_6$, $Rb_2TeCuBr_6$, and $Cs_2NiBaI_6$ occupy the narrower-gap region and are more relevant for absorber-oriented discussion, whereas $K_2MnCdCl_6$ and $Cs_2AgInCl_6$ serve better as stable semiconducting reference systems. By contrast, $Rb_2FeMnF_6$ and $Cs_2GeSrBr_6$ remain structurally valid but electronically too wide-gap for strong absorber candidacy. The DFT step provides practical thermodynamic closure by confirming that the shortlisted compounds remain recognizable members of the ordered $A_2BB'X_6$ family after full relaxation.

Table 8. Final lead-free DFT-validated candidates retained for detailed discussion.

| Serial | Compound | ML-predicted $E_g$ (eV) | DFT $E_g$ (eV) | t | μ | τ | Space group | a (Å) |
|---|---|---|---|---|---|---|---|---|
| 5 | $K_2BePdF_6$ | 1.556 | 0.876 | 0.965 | 0.492 | 3.858 | P1 | 10.340 |
| 8 | $K_2MnCdCl_6$ | 1.381 | 1.060 | 0.861 | 0.448 | 4.432 | P1 | 10.914 |
| 41 | $Rb_2TeCuBr_6$ | 1.289 | 0.742 | 0.820 | 0.531 | 4.736 | P1 | 11.488 |
| 44 | $Cs_2SnGeBr_6$ | 1.167 | 1.112 | 0.881 | 0.487 | 4.174 | P1 | 12.000 |
| 46 | $Cs_2GeSrBr_6$ | 1.584 | 3.434 | 0.881 | 0.487 | 4.181 | P1 | 12.100 |
| 61 | $Cs_2NiBaI_6$ | 1.330 | 0.656 | 0.850 | 0.464 | 4.478 | P1 | 12.063 |
| 114 | $Cs_2AgInCl_6$ | 1.268 | 1.419 | 0.846 | 0.608 | 4.282 | P1 | 11.300 |

*Only Pb-free compounds are retained in the final validated set shown in the main text.

3.7 Structural assignability and simulated XRD fingerprints

After DFT validation, simulated powder X-ray diffraction patterns were generated from the relaxed structures to evaluate the structural assignability of the final Pb-free $A_2BB'X_6$candidates. This step is important because the backward-mapping workflow identifies promising compositions computationally, but experimental translation requires phase-specific diffraction fingerprints that can be compared with future powder XRD measurements. Therefore, the simulated XRD patterns are used here as crystallographic fingerprints of the relaxed DFT models, rather than as experimental proof of phase formation.

The simulated diffraction profiles show that the final candidates have distinguishable peak distributions, even when they belong to related halide families. Most major reflections appear in the $15°–45°$ $2\theta$range, with composition-dependent shifts that reflect differences in lattice parameter, halide size, and B/B′-site chemistry. This indicates that the shortlisted compounds are not only chemically distinct, but also structurally traceable through powder diffraction.

The fluoride and chloride candidates show clear diffraction patterns. $K_2BePdF_6$ has a strong simulated reflection near $2\theta = 24.35°$, while $K_2MnCdCl_6$ and $Cs_2AgInCl_6$ show their strongest peaks near 32.80° and 22.25°, respectively (see Figure S5 and Table S9). These distinct reflections suggest that fluoride and chloride candidates should be easy to differentiate in future phase-identification experiments. Bromide and iodide candidates have diffraction patterns that are closer together but can still be distinguished by secondary reflections and relative intensity patterns. $Rb_2TeCuBr_6$ has its strongest reflection near 31.10°, while $Cs_2SnGeBr_6$, $Cs_2GeSrBr_6$, and $Cs_2NiBaI_6$ show dominant peaks around 29.5°–29.75° (see Figure S5 and Table S9). This clustering is typical of larger metal–halide frameworks. However, additional reflections between 20.7°–25.7° and 42.2°–42.6° provide unique phase-specific fingerprints for further distinction.

These simulated XRD patterns confirm that the DFT-relaxed structural models produce assignable diffraction fingerprints, but experimental powder XRD, impurity-phase comparison, and Rietveld refinement would still be required to establish phase purity under synthesis conditions. Within the present computational workflow, the XRD analysis adds a practical structural layer by demonstrating that the final candidates are experimentally testable. The full set of relaxed crystal structures and simulated XRD patterns is provided in Figure S5.

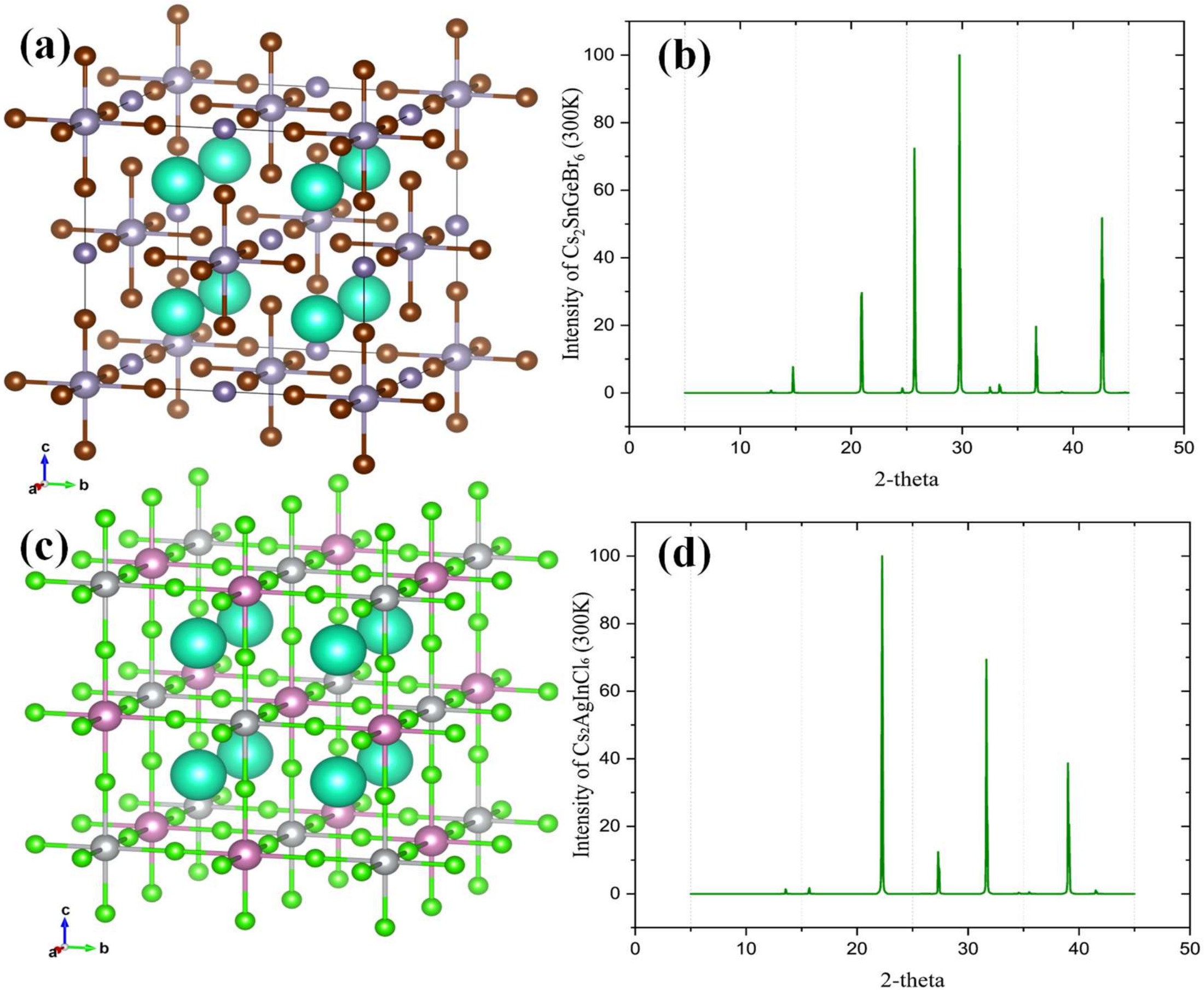

Figure 10. Optimized crystal structures and simulated powder XRD patterns of representative DFT-validated Pb-free $A_2BB'X_6$candidates: (a,b) $Cs_2SnGeBr_6$, and (c,d) $Cs_2AgInCl_6$.

### 3.8 Band-edge orbital hybridization and effective-mass-limited carrier transport

Band structures and PDOS (see Figure S6) were used to examine band-edge electronic structure and intrinsic transport tendency. Electron and hole effective masses were extracted from the curvature of the conduction-band minimum (CBM) and valence-band maximum (VBM), respectively. Since carrier mobility is inversely related to effective mass under comparable scattering conditions, these values are discussed as effective-mass-limited transport descriptors. As summarized in Table S10, the investigated compounds exhibit band gaps from 0.656 eV ($Cs_2NiBaI_6$) to 3.434 eV ($Cs_2GeSrBr_6$), indicating broad electronic tunability. Most systems show direct or nearly direct band gaps, whereas $Cs_2NiBaI_6$ exhibits an indirect C - Γ transition. Direct-gap compounds are generally better for optoelectronic processes. These processes involve radiative transitions.

Effective masses vary significantly. This reflects differences in band-edge curvature and orbital composition. Dispersive s- and p-orbital-derived band edges result in low carrier effective masses, whereas localized transition-metal d-dominated valence states produce flatter valence bands and heavier holes (see Figure S6 and Table S10). This trend is evident for $K_2MnCdCl_6$ and $Cs_2AgInCl_6$, where metal-d/halogen-p-dominated VBMs lead to extremely large hole masses of 170.6 $m_0$ and 84.6 $m_0$, respectively, indicating strong hole localization and electron-dominant transport. $Cs_2SnGeBr_6$ shows the most favorable hole-transport descriptor, with an ultralow $m_h^*$ of 0.054 $m_0$ , originating from the highly dispersive Br-p/Sn–Ge-hybridized VBM (see Figure S6 and Table S10). Combined with a low electron mass of 0.26 $m_0$ and a direct band gap of 1.112 eV, this compound is expected to support efficient carrier generation and transport. $Rb_2TeCuBr_6$ exhibits attractive ambipolar behavior with balanced electron and hole masses of 0.32 $m_0$ and 0.44 $m_0$, respectively; its Cu-d/Br-p-derived VBM and Te/Br-derived CBM provide sufficient dispersion at both band edges, beneficial for carrier collection in photovoltaic and photodetector devices (see Figure S6 and Table S10).

$Cs_2AgInCl_6$ shows highly asymmetric transport. Its dispersive In-s/Cl-p-derived conduction-band minimum (CBM) gives the lowest electron mass in the series (0.078 $m_0$). Meanwhile, the flat Ag-d/Cl-p-dominated valence-band maximum (VBM) leads to a very heavy hole mass. Similarly, $K_2MnCdCl_6$ has a light electron mass and an extremely heavy hole mass. This suggests that both materials are better for electron-dominant transport than for balanced ambipolar conduction.

$Cs_2NiBaI_6$ exhibits a light hole mass of 0.13 $m_0$ suggesting favorable p-type transport, but its indirect band gap may reduce radiative transition efficiency. $Cs_2GeSrBr_6$ shows very light electrons due to a dispersive CBM, but its wide band gap of 3.434 eV makes it more suitable for UV or transparent optoelectronic applications. $K_2BePdF_6$ displays moderate electron and hole masses, indicating reasonable but less competitive transport.

The combined band-structure and PDOS analysis demonstrates that band-edge orbital hybridization is a key descriptor for carrier transport (Figure S6). Compounds with dispersive s/p-orbital-derived band edges, particularly $Cs_2SnGeBr_6$ and $Rb_2TeCuBr_6$, exhibit low effective masses and favorable ambipolar transport tendencies, whereas localized transition-metal d-dominated VBMs lead to flat bands and hole localization.

### 3.10 Optical phenotype of the shortlisted compounds

Optical-response analysis is the decisive step for distinguishing merely stable semiconductors from candidates plausibly useful for absorber-centered optoelectronics. The phenotype suite derived from the complex dielectric function is organized as a unified discussion of dielectric screening, near-edge absorption, optical impedance, loss spectra, and optical conductivity. A strong inverse-design claim requires validation beyond band-gap prediction; we therefore evaluate a complete optical phenotype suite where $\alpha(\omega)$ measures light harvesting, $n(\omega)$ and $R(\omega)$ describe optical impedance, $L(\omega)$ fingerprints excitation regimes, and $\sigma(\omega)$ quantifies transition strength.

### 3.10.1 Dielectric-function analysis and optical screening behavior

The complex dielectric function $\varepsilon(\omega) = \varepsilon_1(\omega) + i\varepsilon_2(\omega)$ was evaluated to understand optical response, electronic polarizability, and interband transition characteristics. $\varepsilon_1(\omega)$ describes dispersive response and electronic screening capability, while $\varepsilon_2(\omega)$ represents absorptive response associated with interband transitions. The static dielectric constant $\varepsilon_1(0)$, zero-frequency refractive index $n(0) = \sqrt{\varepsilon_1(0)}$, and the dominant $\varepsilon_2(\omega)$ peaks were extracted and summarized (see Table S11 and Figure S7). The investigated double perovskites exhibit distinct dielectric responses reflecting differences in band gap, orbital composition, and electronic polarizability.

The static dielectric constant varies a lot. It goes from 3.46 in $Cs_2AgInCl_6$ to 15.28 in $Cs_2NiBaI_6$ (see Table S11 and Figure S7). This shows that low-energy electronic screening depends on chemical composition. Compounds with heavier halides are more polarizable. They also have narrower band gaps. These compounds tend to have larger dielectric constants. $Cs_2NiBaI_6$ has the largest dielectric constant. It is $\varepsilon_1(0) = 15.28$ and has a refractive index of $n(0) = 3.91$. This means it has strong electronic polarizability. It efficiently screens Coulombic interactions. The strongest $\varepsilon_2$ peak is at 6.02 eV. This shows that interband transitions happen at high photon energies, even though it has a narrow band gap.

$Rb_2TeCuBr_6$ has one of the best optical responses. It has a static dielectric constant of 13.27. Its strongest absorptive dielectric peak is $\varepsilon_2^{max} = 10.45$, at 1.30 eV (see Table S11). This shows strong interband transitions in the visible/near-infrared region. With its narrow band gap and balanced effective masses, $Rb_2TeCuBr_6$ is promising for photovoltaic and photodetector applications.

$Cs_2SnGeBr_6$ has a balanced dielectric profile. Its $\varepsilon_1(0) = 7.54$ and $n(0) = 2.75$. The strong $\varepsilon_2(\omega)$ peak is at 3.06 eV (Table S11). This shows good electronic screening. The interband transitions extend into the visible-to-near-UV region.

$K_2BePdF_6$ also has a high static dielectric response. Its $\varepsilon_1(0) = 11.27$ and $n(0) = 3.36$. The strongest $\varepsilon_2(\omega)$ feature occurs at 0.37 eV (Table S11). This suggests strong electronic polarizability. It may also have near-infrared optical activity. However, its moderate effective masses suggest less favorable carrier transport.

In contrast, $Cs_2GeSrBr_6$ and $Cs_2AgInCl_6$ exhibit comparatively weak static screening, with $\varepsilon_1(0)$ values of 3.71 and 3.46, respectively. Their dominant $\varepsilon_2(\omega)$peaks occur at higher photon energies, consistent with a UV-dominant optical response. For $Cs_2GeSrBr_6$, the wide band gap of 3.434 eV and the main absorptive feature at 4.23 eV suggest potential relevance for ultraviolet or transparent optoelectronic applications. Similarly, $Cs_2AgInCl_6$ shows its main absorptive response at 14.02 eV, indicating limited visible-light absorption despite its favorable electron effective mass (Table S11). Overall, the dielectric-function analysis confirms that optical screening and interband transition strength are strongly composition-dependent. Narrow-gap, highly polarizable systems exhibit enhanced static dielectric constants. Among them, $Rb_2TeCuBr_6$ is especially notable because it combines strong low-energy absorption, high dielectric screening, and balanced carrier transport.

3.10.2 Absorption-coefficient response and target-driven light-harvesting assessment

The absorption coefficient α(ω) was examined to evaluate light-harvesting capability (see Table S12 and Figure S8). The absorption edge was estimated as the first photon-energy region where α reaches approximately $10^4$ $cm^{-1}$. The visible–NIR maximum was extracted from the low-energy optical window, while the UV maximum was obtained from the higher-energy region. The DFT-validated compounds display strongly composition-dependent absorption behavior, demonstrating that band gap alone is insufficient to determine optical device suitability. Absorption strength depends on the combined effects of band-edge orbital character, transition probability, dielectric response, and band dispersion.

$Rb_2TeCuBr_6$ shows one of the most favorable low-energy absorption profiles, reaching a visible–NIR absorption maximum of $0.559 \times 10^5$ $cm^{-1}$ at 1.93 eV capability (see Table S12 and Figure S8). This efficient photon absorption in a technologically relevant spectral region is consistent with its narrow band gap, strong dielectric response, and balanced effective masses. From a chemical-genome perspective, this favorable response is associated with the polarizable bromide framework and Cu/Te-derived band-edge states supporting strong interband transitions. $Cs_2SnGeBr_6$ also satisfies important optoelectronic target criteria, with a strong visible-to-near-UV absorption maximum of $0.642 \times 10^5$ $cm^{-1}$ near 3.08 eV and a larger UV response of $0.802 \times 10^5$ $cm^{-1}$ capability (see Table S12 and Figure S8). The Sn/Ge–Br motif provides a useful design rule: dispersive s/p-orbital-derived band edges combined with a polarizable halide sublattice produce a balanced optoelectronic response (see Table 9 and 10).

By contrast, $Cs_2GeSrBr_6$ and $Cs_2AgInCl_6$ are better aligned with UV-active or transparent targets. $Cs_2GeSrBr_6$ has weak visible-region absorption with stronger UV response, while $Cs_2AgInCl_6$ shows weak visible absorption despite its very light electron effective mass. This highlights that favorable charge transport alone is insufficient for photovoltaic performance if optical transition strength in the target spectral region is weak. $Cs_2NiBaI_6$ shows why multi-descriptor screening is important. It has the smallest band gap, 0.656 eV. Its strongest absorption happens in the UV region. This is not near the band edge. It is likely due to indirect-gap character. Band-edge orbital selection effects also play a role. $Rb_2TeCuBr_6$ and $Cs_2SnGeBr_6$ meet the combined optical and transport requirements. They are the best for high-performance lead-free optoelectronic absorbers.

3.10.3 Refractive-index response and optical dispersion

The complex refractive index $\tilde{n}(\omega) = n(\omega) + i\kappa(\omega)$ was used. This helps study optical dispersion and attenuation (see Table S13 and Figure S9). The zero-frequency refractive index n(0) follows the same trend as $\varepsilon_1(0)$. $Cs_2NiBaI_6$ has the largest n(0) = 3.94. $Rb_2TeCuBr_6$ follows with n(0) = 3.64. $K_2BePdF_6$ has n(0) = 3.38.

$Rb_2TeCuBr_6$ has the best refractive-index profile for low-energy optoelectronic applications. It has a high n(0). It also has the largest visible–NIR extinction coefficient (k_max = 1.937 at 1.63 eV).

$Cs_2SnGeBr_6$ has a well-balanced response. Its n(ω) reaches 2.79 at 0.92 eV. κ increases strongly near the visible-to-near-UV region (κ = 1.291 at 3.08 eV) (see Table S13).

$Cs_2GeSrBr_6$ and $Cs_2AgInCl_6$ have low static refractive indices. $Cs_2GeSrBr_6$ has n(0) = 1.93. $Cs_2AgInCl_6$ has n(0) = 1.86. These compounds show weak visible–NIR extinction. Their absorption is UV-dominant. $Cs_2NiBaI_6$ has a large n(0). It also has significant low-energy extinction. However, its indirect band gap and weaker near-edge absorption show that a high refractive index alone does not make it a good photovoltaic absorber.

3.10.4 Reflectivity response and optical-loss assessment

The reflectivity spectrum R(ω) was analyzed to evaluate optical reflection losses (see Table S14 and Figure S10). $Rb_2TeCuBr_6$ shows the highest visible–NIR reflectivity ($R_{max} = 0.373$at 1.53 ), consistent with its high dielectric constant and strong low-energy absorption. This indicates strong light–matter interaction but also suggests that surface-reflection losses should be minimized through antireflection coatings or surface texturing. $Cs_2SnGeBr_6$ has moderate-to-high reflectivity. Its R_max is 0.259 near 3.08 eV. It has lower visible–NIR reflectivity than $Rb_2TeCuBr_6$. It still maintains strong absorption. It has favorable carrier effective masses.

$Cs_2NiBaI_6$ has the largest static reflectivity. Its R(0) is 0.360 (see Table S14). This matches its large $\varepsilon_1(0)$ and n(0). However, it has weak near-edge absorption. High polarizability alone does not make it an optimal photovoltaic absorber. $Cs_2AgInCl_6$ and $Cs_2GeSrBr_6$ have low static reflectivity. $Cs_2AgInCl_6$ has R(0) = 0.131. $Cs_2GeSrBr_6$ has R(0) = 0.100. Low reflectivity is useful only with strong absorption in the target spectral range.

3.10.5 Energy-loss function and plasmonic optical-loss response

The electron energy-loss function L(ω) was calculated to evaluate optical dissipation (see Table S15 and Figure S11). $Cs_2SnGeBr_6$ shows the most favorable loss behavior for absorber-oriented applications, with comparatively low visible-region loss ($L_{max} = 0.147$ near 3.18 eV) (see Table S15) combined with a direct band gap, strong absorption, and low effective masses. $Rb_2TeCuBr_6$ shows moderate visible-region loss. Its L_max is 0.371 at 2.60 eV. It also has a strong high-energy loss peak. This suggests that reflection and dissipative losses should be considered during device optimization. $Cs_2NiBaI_6$ has the strongest high-energy loss peak. This matches its large dielectric constant. However, its indirect band gap and weak near-edge absorption show that high polarizability alone is not enough. $Cs_2GeSrBr_6$ shows very weak visible-region loss. This matches its wide band gap and limited visible absorption. This may be useful for transparent or UV-selective applications.

The energy-loss analysis supports the optical design hierarchy. This hierarchy comes from the dielectric-function, refractive-index, reflectivity, absorption, and effective-mass results. $Cs_2SnGeBr_6$ and $Rb_2TeCuBr_6$ are the strongest absorber-oriented candidates (see Table S15). They have different design trade-offs. $Cs_2SnGeBr_6$ combines strong absorption with comparatively low visible-region loss, whereas $Rb_2TeCuBr_6$ provides stronger low-energy light–matter interaction but may require more careful optical-loss management. These results reinforce the main concept of this work: phase-stable lead-free double perovskites should be discovered through backward mapping from device targets to chemical genomes, where absorption strength, dielectric screening, reflectivity loss, plasmonic dissipation, band-edge orbital character, and carrier transport are optimized together rather than treated as isolated descriptors.

### 3.10.6 Optical-conductivity response and photoinduced charge-excitation behavior

The frequency-dependent optical conductivity, $\sigma(\omega) = \sigma_1(\omega) + i\sigma_2(\omega)$, was analyzed to further evaluate the photoinduced charge-excitation behavior of the DFT-validated lead-free double perovskites. The real component, $\sigma_1(\omega)$, represents the dissipative optical conductivity associated with interband electronic transitions, whereas the imaginary component, $\sigma_2(\omega)$, describes the reactive polarization response of the material. In the context of backward mapping from device targets to chemical genomes, optical conductivity is a useful descriptor because it indicates how effectively a material converts incident photons into electronic excitations within a target spectral window. Therefore, $\sigma(\omega)$provides a complementary bridge between optical absorption, dielectric screening, and band-edge transport behavior. The extracted optical-conductivity descriptors are summarized in Table S16, and the corresponding spectra are shown in Figure S12.

As shown in Figure S12 and summarized in Table S16, the optical-conductivity spectra exhibit strong composition dependence across the DFT-validated chemical space. The most intense $\sigma_1(\omega)$peaks generally occur at higher photon energies, indicating that the strongest interband charge excitations are located in the UV or high-energy region for several compounds. This trend is consistent with the dielectric-function, absorption, and energy-loss analyses, where high-energy optical transitions dominate in wider-gap or weakly visible-active materials. These results emphasize that a favorable electronic band gap alone is not sufficient to define an efficient optoelectronic absorber; the optical transition strength within the target spectral region must also be considered.

Among the investigated compounds, $Rb_2TeCuBr_6$ shows the strongest low-energy charge-excitation response, with a visible–NIR $\sigma_1$maximum of 1.801 at 1.53 eV (see Table S16). This feature agrees well with its strong absorption coefficient, high dielectric response, large refractive index, and balanced electron–hole effective masses. From the backward-mapping perspective, $Rb_2TeCuBr_6$ therefore satisfies multiple absorber-oriented device targets simultaneously, including efficient low-energy photon-to-charge excitation, strong light–matter interaction, and favorable effective-mass-limited carrier transport. This makes the Rb–Te–Cu–Br chemical motif a particularly interpretable design direction for low-energy optoelectronic absorbers. $Cs_2SnGeBr_6$ also exhibits a highly favorable optical-conductivity profile. Its visible-to-near-UV conductivity reaches 2.129 at 3.18 eV, while the global maximum occurs at 13.80 eV (see Table S16). Importantly, this strong conductivity response is accompanied by a direct band gap, high absorption coefficient, relatively low visible-region energy loss, and low carrier effective masses. These combined properties suggest that $Cs_2SnGeBr_6$ can support efficient optical excitation and subsequent charge transport, making it another promising absorber-oriented candidate.

Chemically, the Sn/Ge–Br framework appears to promote dispersive band-edge states and strong optical transitions, providing a useful design rule for lead-free double-perovskite discovery.

In contrast, $Cs_2GeSrBr_6$ and $Cs_2AgInCl_6$ display weaker visible-region optical conductivity but pronounced high-energy conductivity peaks. $Cs_2GeSrBr_6$ reaches its strongest response at 14.00 eV, while $Cs_2AgInCl_6$ shows the largest global conductivity among the series at 14.08 eV (Table S16). This behavior is consistent with their UV-dominant absorption and loss-function responses. Consequently, these compositions are better mapped to UV-active or transparent optoelectronic targets than to broadband photovoltaic absorbers, despite the favorable electron effective mass observed for $Cs_2AgInCl_6$. $Cs_2NiBaI_6$ illustrates why multi-descriptor validation is essential in target-driven materials discovery. Although it has the smallest electronic band gap in the series, its visible-region optical conductivity is relatively weak, while its strongest conductivity feature appears at 6.27 eV (see Table S16). This indicates that its near-edge optical transitions are not sufficiently strong for efficient visible-light harvesting, likely due to its indirect-gap character and band-edge orbital selection effects. Therefore, $Cs_2NiBaI_6$ should not be prioritized solely on the basis of its narrow gap, even though it exhibits strong dielectric screening and a light hole effective mass.

Overall, the optical-conductivity analysis reinforces the design hierarchy obtained from the absorption coefficient, dielectric function, refractive index, reflectivity, energy-loss spectra, and carrier effective masses. $Rb_2TeCuBr_6$ and $Cs_2SnGeBr_6$ remain the most promising absorber-oriented candidates because they combine strong optical conductivity in relevant energy windows with favorable transport and screening descriptors. By contrast, $Cs_2GeSrBr_6$ and $Cs_2AgInCl_6$ are more naturally suited to UV-active or transparent optoelectronic applications. These findings support the central premise of this work: reliable discovery of phase-stable lead-free double perovskites requires backward mapping from device targets to chemical genomes using a coupled descriptor set that includes optical conductivity, absorption strength, dielectric screening, reflection loss, plasmonic dissipation, band-edge orbital character, and carrier transport.

### 3.11 Chemical-genome design rules revealed by backward mapping

The purpose of the backward-mapping workflow is not only to identify promising lead-free double perovskites, but to understand why specific chemistries outperform others. Here, the final DFT-validated candidates are interpreted through the same chemical-genome framework used for descriptor construction, ML screening, and candidate down-selection. Input descriptors are treated as chemical genes, while DFT-calculated structural, electronic, dielectric, optical, and transport properties are treated as observable phenotypes. The six-family organization is essential because device performance emerges from coupled descriptor contributions: a useful candidate must be structurally feasible, chemically cohesive, electronically well aligned, optically active, sufficiently screened, and transport-compatible.

Here, Figure 11 summarizes the multi-phenotype performance of the DFT-validated candidates. The heat map shows that no compound is optimal across all properties. $Rb_2TeCuBr_6$ and $Cs_2SnGeBr_6$ show the most balanced absorber-oriented profiles, whereas $Cs_2GeSrBr_6$ and $Cs_2AgInCl_6$ are more naturally suited to UV-active, transparent, or electron-selective optoelectronic roles. Figure 12 translates these trends into a chemical-genome design-rule decoder, linking each descriptor family to its corresponding DFT phenotype and device-level implication.

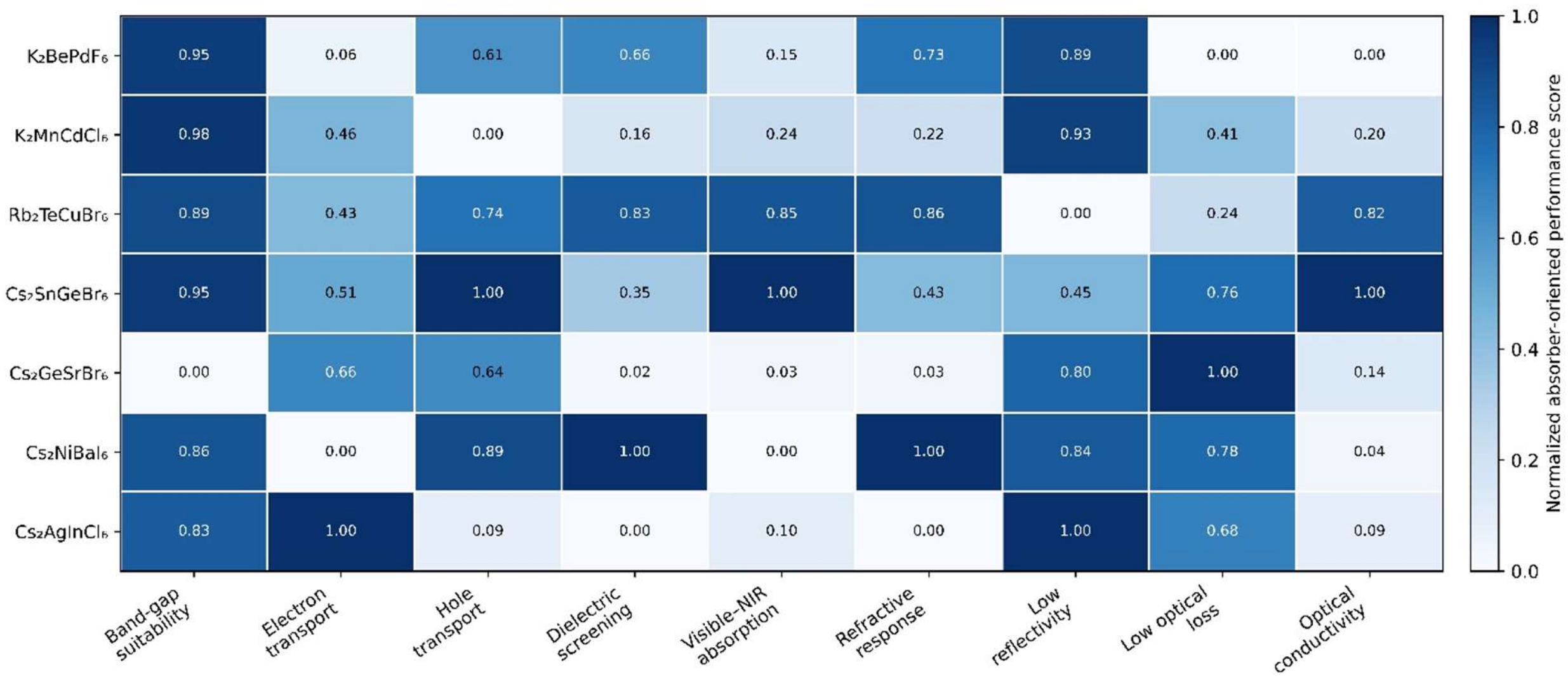

Figure 11. Multi-phenotype performance map of DFT-validated lead-free double perovskites. Normalized heat map comparing absorber-relevant electronic, transport, dielectric, and optical phenotypes of the DFT-validated candidates. Darker blue indicates a more favorable absorber-oriented score. $Rb_2TeCuBr_6$ and $Cs_2SnGeBr_6$ show the most balanced absorber profiles, while $Cs_2GeSrBr_6$ and $Cs_2AgInCl_6$ are more naturally suited to UV-active, transparent, or electron-selective optoelectronic targets.

### 3.11.1 Geometric packing genes define the structural feasibility window

The first requirement in the design sequence is geometric feasibility. Packing descriptors, including ionic radii, the Goldschmidt tolerance factor $t$, the octahedral factor $\mu$, and the new tolerance factor $\tau$, determine whether the selected A-site cation, B/B′-site pair, and halide anion can be accommodated within the $A_2BB'X_6$framework. These descriptors are often used as simple formability filters, but the present results show that their role is broader. Even inside the formable region, size matching affects octahedral distortion, metal-halide bond geometry, orbital overlap, and band dispersion near the band edges.

The practical rule is therefore to enforce packing feasibility before optimizing optical or transport properties (See Figure 12, Table 9 and 10). A composition outside the formability manifold is unlikely to remain structurally meaningful after relaxation, even if its predicted band gap appears attractive. At the same time, satisfying the packing condition alone is not enough. Several DFT-validated candidates are geometrically feasible and semiconducting, but they differ strongly in absorption strength, dielectric response, carrier effective mass, and optical loss. Packing genes therefore define the structural search region, while the remaining gene families determine whether that structure becomes useful for a specific device target (See Figure 12, Table 9 and 10).

### 3.11.2 Framework cohesion genes connect stability with metal–halide robustness

Once geometric feasibility is satisfied, framework cohesion becomes the next control layer. Bond-energy proxies, metal–halide bond strength, and formation-energy-related descriptors determine whether the $BX_6$ and $B'X_6$ octahedral network is chemically robust. These descriptors are directly connected to thermodynamic accessibility, but they also influence electronic and optical behavior because metal–halide bonding controls orbital hybridization at the band edges.

This connection is important because a suitable band gap alone does not guarantee useful absorption.

$Cs_2NiBaI_6$ has a narrow band gap and strong dielectric screening. However, its near-edge optical response is weaker than expected for a primary visible-light absorber. This suggests that the lowest-energy transition is not strongly optically active. It is likely due to indirect-gap character and band-edge selection effects.

In contrast, $Rb_2TeCuBr_6$ and $Cs_2SnGeBr_6$ show stronger visible-to-near-UV optical responses. Their framework chemistry supports more effective interband transitions.

The design rule is therefore to optimize cohesion inside the formable region, not in isolation. Strong B–X and B′–X interactions are useful when they both stabilize the framework and generate band-edge states that support optically allowed transitions (See Figure 12,Table 9 and 10). In this sense, framework cohesion genes act as the bridge between phase stability and optical functionality.

### 3.11.3 Chemical polarity and covalency genes tune band-edge hybridization

Chemical polarity and covalency descriptors, including electronegativity differences and metal–halide chemical contrast, provide a second bonding-related control layer.

Framework cohesion describes the chemical robustness of the lattice. Polarity/covalency genes describe how strongly the metal and halide orbitals mix. This distinction is important. Optical transition strength depends on the size of the band gap. It also depends on the orbital character and symmetry of the valence- and conduction-band edges.

In the DFT-validated set, the strongest absorber-oriented candidates are those in which the bonding environment supports both stability and optically active band-edge hybridization. $Rb_2TeCuBr_6$ and $Cs_2SnGeBr_6$ are favorable because their optical response is not only a consequence of band-gap placement, but also of band-edge chemistry, absorption coefficient, and optical conductivity. By contrast, candidates with weaker near-edge transition strength or less favorable orbital coupling are better assigned to specialized targets rather than general photovoltaic absorption.

The practical rule is to tune metal–halide covalency so that it strengthens useful band-edge transitions without destabilizing the structure (See Figure 12, Table 9 and 10). Electronegativity contrast should therefore be optimized together with cohesion and band dispersion, rather than treated as an isolated descriptor.

### 3.11.4 Charge-transfer genes regulate chemical alignment and stability tendency

Charge-transfer genes include ionization energy, electron affinity, and formation-energy balance across the A, B/B′, and X sites. These descriptors influence how easily charge can be redistributed within the lattice and whether the chosen elements support chemically reasonable oxidation states. They also affect band alignment because the energy positions of the valence and conduction states depend on the redox character of the constituent ions.

In the present workflow, charge-transfer descriptors help explain why some compositions that appear promising from a band-gap perspective still require DFT validation before device assignment. A compound may fall within the target gap window but remain chemically fragile if its elemental chemistry favors competing charge states or decomposition products. Charge-transfer genes therefore act as a chemical-plausibility layer between geometric screening and final electronic-structure interpretation.

The design rule is to avoid compositions where the target electronic response relies on fragile charge balance. Stable charge accommodation should be considered before assigning a material to absorber, UV-active, transparent, or transport-specific roles (See Figure 12, Table 9 and 10).

### 3.11.5 Polarization and screening genes control dielectric and refractive response

Polarization and screening genes connect elemental polarizability, halide softness, ionization energy, and electron affinity to dielectric response and refractive behavior. These descriptors are important for optoelectronic materials. Stronger dielectric screening can reduce electron–hole Coulomb attraction. It can also lower excitonic penalties and support charge separation. In general, Br- and I-containing frameworks have stronger polarizability than F- or Cl-based analogues. This is consistent with the strong dielectric and refractive responses seen in $Rb_2TeCuBr_6$ and $Cs_2NiBaI_6$.

However, high screening alone does not define an ideal absorber. $Cs_2NiBaI_6$ shows this clearly. It has a narrow band gap and strong dielectric screening. But its indirect-gap character and weak near-edge absorption make it less suitable as a primary visible-light absorber. Therefore, polarization genes must be interpreted together with optical transition strength, optical conductivity, and transport behavior.

The design rule is to use polarizable frameworks to enhance screening only when the same chemistry also supports absorption in the target spectral region. High dielectric response is valuable, but it must be coupled with useful oscillator strength and manageable optical loss (See Figure12, Table 9 and 10).

### 3.11.6 Electronic-identity genes control band-edge character and carrier transport

Electronic-identity genes include valence electron count, atomic number, and B/B′-site orbital character. These descriptors determine which orbitals form the valence-band maximum and conduction-band minimum. They control band dispersion, carrier effective mass, and transport asymmetry.

The DFT band structures and PDOS results show that dispersive s/p-derived band edges generally produce lower effective masses, whereas localized d-dominated valence states tend to produce flat bands and heavy holes (See FigureS6 and Table S10).

This behavior is evident in $K_2MnCdCl_6$ and $Cs_2AgInCl_6$ (See FigureS6 and Table S10). Both compounds show highly localized valence-band character and poor hole-transport tendency. $Cs_2AgInCl_6$ is especially instructive because it combines an ultralight electron branch with a very heavy hole branch. This makes it more suitable for UV-active or electron-selective applications than for balanced photovoltaic absorption.

In contrast, $Rb_2TeCuBr_6$ and $Cs_2SnGeBr_6$ show more favorable band-edge dispersion. $Rb_2TeCuBr_6$ provides relatively balanced electron and hole transport, while $Cs_2SnGeBr_6$ combines favorable carrier masses with strong optical response (See Figure S6 and Table S10). From Table 9 and 10, we get corresponding rule is to favor compositions with delocalized s/p-derived band edges and to avoid strongly localized d-dominated valence states when ambipolar transport is required (See Figure 12).

3.11.7 Optical-loss phenotypes refine absorber assignment

Optical loss is treated here as a phenotype-level check rather than a separate chemical-gene family. Absorption coefficient and optical conductivity describe how efficiently photons generate electronic excitations. Reflectivity, extinction coefficient, and the energy-loss function indicate how much incident energy may be lost through surface reflection or dissipative response. The best absorber is not necessarily the material with the largest absorption peak.

$Rb_2TeCuBr_6$ shows strong visible-near-infrared absorption, high dielectric response, strong optical conductivity, and relatively balanced carrier masses, making it a strong absorber-oriented candidate (see Figure S8, S10 and Table S12, S16). However, its higher visible-NIR reflectivity suggests that optical-stack engineering may be useful in a practical device. $Cs_2SnGeBr_6$ provides a more balanced absorber profile because it combines strong visible-to-near-UV response with favorable transport and comparatively lower visible-region optical loss. In contrast, $Cs_2GeSrBr_6$ and $Cs_2AgInCl_6$ are better interpreted as UV-active, transparent, or electron-selective candidates rather than primary visible-light absorbers.

The device-level rule is that absorber selection should be based on coupled performance, not on a single metric. A strong absorber profile requires formability, stability, optically active band edges, dielectric screening, balanced transport, and manageable optical loss (See Figure 12, Table 9 and 10).

Table 9. Descriptor-phenotype coupling from the six-family chemical-genome framework.

| Chemical-gene family | Representative descriptors | Main DFT phenotype | Design interpretation |
|---|---|---|---|

| Geometric packing genes | Ionic radii; t, μ, τ; A-, B/B′-, and X-site size matching | Structural feasibility, octahedral distortion, lattice relaxation behavior | Defines the formability window in which the $A_2BB'X_6$ framework can exist. Packing is the first structural filter but does not alone determine device performance. |
|---|---|---|---|
| Framework cohesion genes | A–X, B–X, and B′–X bond-energy proxies; formation-energy-related descriptors | Phase stability, octahedral robustness, metal–halide framework persistence | Controls whether the relaxed structure is chemically robust enough for further functional optimization. |
| Chemical polarity/ covalency genes | Electronegativity differences; X–B, X–B′, and X–X descriptors | Band-edge hybridization, optical transition strength, $\varepsilon_2(\omega)$ | Tunes the orbital mixing that controls near-edge optical activity and absorption strength. |
| Charge-transfer genes | Ionization energy, electron affinity, formation-energy balance | Charge accommodation, redox plausibility, band alignment | Helps avoid chemically fragile compositions and supports stable electronic alignment. |
| Polarization/ screening genes | Atomic polarizability, halide polarizability, ionization energy/electron-affinity response | Static dielectric constant, refractive index, excitonic screening, low-energy optical response | Enhances screening and light–matter interaction but must be paired with optically allowed transitions. |
| Electronic-identity genes | Valence electron count, atomic number, B/B′-site orbital character | PDOS, band dispersion, effective mass, carrier asymmetry | Determines whether band edges are dispersive s/p states or localized d states, controlling transport suitability. |

Table 10. DFT-validated chemical-genome design rules for future A2 BB'X6 discovery.

| Design rule | Chemical meaning | Actionable inverse-design lever | Expected device impact |
|---|---|---|---|
| Enforce formability | Ionic-size matching defines whether the | Apply t, μ, τ, and ionic-radius constraints before | Reduces false positives and improves structural realizability |
| before functionality | double-perovskite framework can form | electronic or optical screening | |
| Optimize cohesion inside the formable region | Metal–halide bonding controls framework stability | Favor balanced B–X and B′–X bonding within the target band-gap region | Improves thermodynamic accessibility and structural robustness |
| Tune polarity/ covalency for optically active band edges | Metal–halide hybridization controls transition strength | Adjust B/B′ chemistry and electronegativity contrast to promote allowed near-edge transitions | Strengthens absorption and optical conductivity |
| Maintain chemically plausible charge transfer | Stable charge accommodation supports realistic oxidation chemistry | Use ionization energy, electron affinity, and formation-energy descriptors to avoid fragile charge balance | Improves chemical stability and band alignment |
| Combine polarizability with oscillator strength | Dielectric screening is useful only when optical transitions are active | Use Br/I polarizability cautiously and validate absorption and optical conductivity | Enhances charge separation without selecting weak-transition false positives |
| Favor dispersive electronic identity for absorbers | s/p-derived band edges reduce effective masses; d-localized VBMs produce heavy holes | Use PDOS and band dispersion to avoid localized valence states in absorber targets | Improves ambipolar transport and reduces carrier imbalance |
| Evaluate optical loss after absorption | Strong absorption can coexist with high reflectivity or energy loss | Use $R(\omega)$, $L(\omega)$, $k(\omega)$, and $\sigma(\omega)$jointly | Improves photon coupling and practical device suitability |
| Match chemistry to natural device role | Not every stable semiconductor is a photovoltaic absorber | Assign candidates to PV, photodetector, UV, transparent, or electron-selective targets based on coupled descriptors | Enables target-specific discovery instead of single-score ranking |

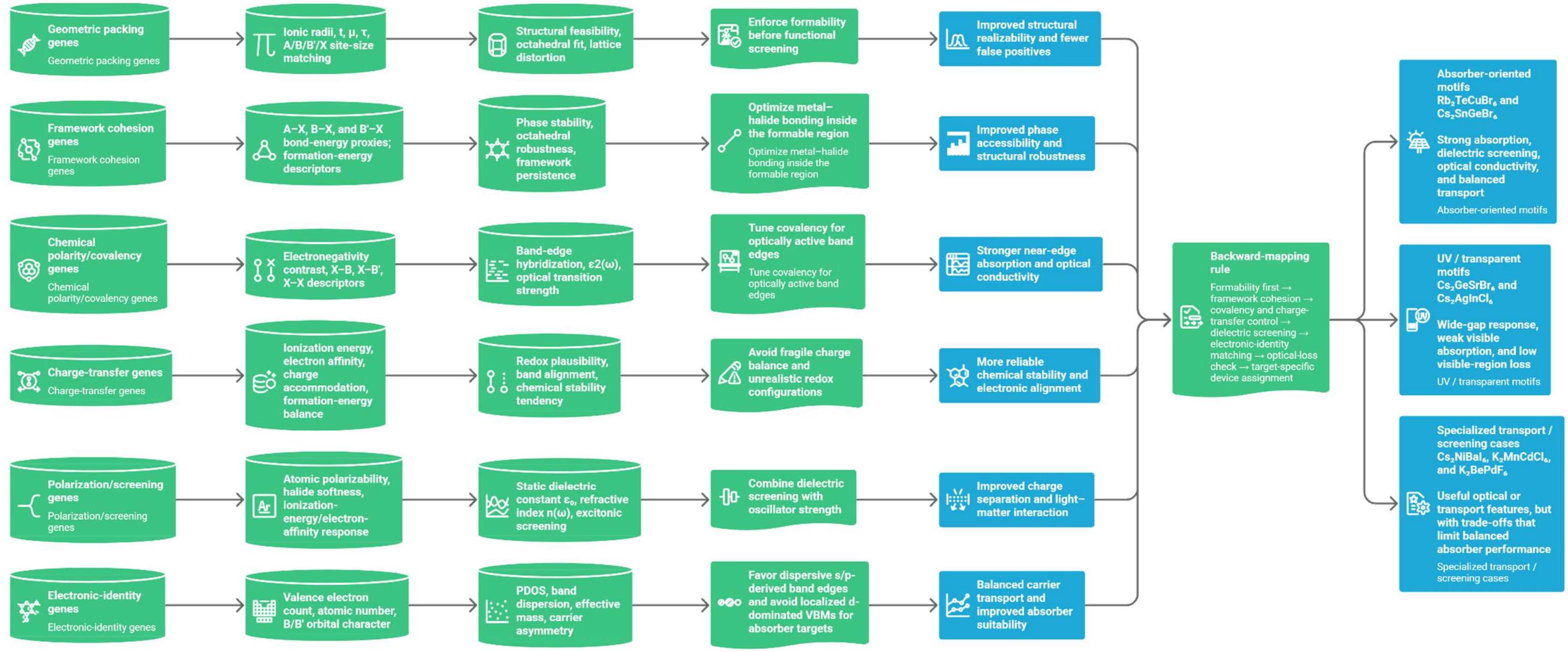

Figure 12. Schematic overview of the chemical-genome design-rule decoder for lead-free double perovskites. The diagram shows the backward-mapping design logic. It links chemical-gene families to DFT phenotypes and device-level outcomes. Six chemical-gene families are mapped to key device attributes. These families are: geometric packing, framework cohesion, chemical polarity/covalency, charge-transfer, polarization/screening, and electronic-identity genes. The attributes include structural feasibility, phase stability, optical-transition strength, dielectric response, band-edge character, transport behavior, and optical-loss constraints. Absorber-oriented candidates must satisfy multiple criteria. These criteria include formability, cohesion, optically active band edges, dielectric screening, balanced transport, and manageable optical loss.

### 3.11.8 Target-specific assignment of the DFT-validated compounds

The backward-mapping framework also allows each validated material to be assigned to the device class it most naturally supports. $Rb_2TeCuBr_6$ is the strongest low-energy absorber candidate because it combines a narrow band gap, strong visible–NIR absorption, high dielectric screening, strong optical conductivity, and relatively balanced carrier masses. Its higher visible–NIR reflectivity suggests that optical-loss management should be considered in device implementation, but its overall descriptor profile remains favorable for photovoltaic and broadband photodetector applications (see Table 11).

$Cs_2SnGeBr_6$ provides the most balanced absorber-type response. It combines strong visible-to-near-UV absorption, favorable carrier masses, direct-gap behavior, and comparatively lower visible-region energy loss. This makes the Cs-Sn-Ge-Br motif a particularly attractive chemical direction for lead-free optoelectronic absorbers.

By contrast, $Cs_2GeSrBr_6$ and $Cs_2AgInCl_6$ are better aligned with UV-active or transparent optoelectronic applications. $Cs_2GeSrBr_6$ has a wide band gap. It has weak visible absorption and low visible optical loss. This makes it more suitable for UV or transparent device targets (see Table 11). $Cs_2AgInCl_6$ shows an ultralow electron effective mass. It also has low static reflectivity. Its very heavy hole mass and UV-dominant optical response limit its usefulness for balanced photovoltaic absorption. $Cs_2NiBaI_6$ has a narrow band gap. It shows strong dielectric screening. Its indirect-gap character and weak near-edge optical response make it less suitable as a primary visible-light absorber (see Table 11).

Table 11. Target-specific assignment of DFT-validated lead-free double perovskites.

| Compound | Dominant strengths | Main limitation | Best-matched device target |
|---|---|---|---|
| $Rb_2TeCuBr_6$ | Strong visible–NIR absorption, high dielectric response, strong optical conductivity, balanced carrier masses | Higher visible–NIR reflectivity and moderate optical loss | Photovoltaic absorber; broadband photodetector |
| $Cs_2SnGeBr_6$ | Balanced absorption, favorable carrier masses, direct-gap behavior, lower visible-region loss | Absorption maximum closer to visible/near-UV boundary | Balanced optoelectronic absorber |
| $K_2BePdF_6$ | Low-gap response, moderate dielectric screening, low-energy optical activity | Weaker absorption and less balanced performance than leading absorbers | Near-IR-sensitive optoelectronic candidate |

| $K_2MnCdCl_6$ | Semiconducting gap and reasonable electron-side response | Heavy-hole transport limitation | Electron-dominant or UV-leaning optical response |
|---|---|---|---|
| $Cs_2GeSrBr_6$ | Wide-gap behavior, low visible optical loss, UV response | Weak visible absorption | Transparent or UV optoelectronics |
| $Cs_2NiBaI_6$ | Narrow gap, strong dielectric screening, light-hole branch | Indirect-gap character, weak near-edge absorption, high reflectivity | Specialized p-type or broadband optical-response candidate |
| $Cs_2AgInCl_6$ | Ultralight electron branch, low static reflectivity | Very heavy hole branch and UV-dominant response | UV-active or electron-selective applications |

Overall, the backward-mapping analysis shows that the most promising lead-free double perovskites are not simply those with the smallest band gaps, largest absorption peaks, or lowest individual effective masses. Useful device assignment emerges from a connected sequence of requirements. These include structural formability, framework cohesion, chemically plausible charge transfer, favorable band-edge hybridization, dielectric screening, manageable optical loss, and balanced carrier transport. From Table12, within the DFT-validated set, $Rb_2TeCuBr_6$ and $Cs_2SnGeBr_6$ best satisfy the absorber-oriented target profile, whereas $Cs_2GeSrBr_6$ and $Cs_2AgInCl_6$ are more naturally suited to UV-active, transparent, or electron-selective applications. These results demonstrate the value of backward mapping from device targets to chemical genomes, where interpretable descriptor families guide the rational discovery of phase-stable lead-free double perovskites with DFT-validated design rules.

## 4. Conclusion

This study developed a backward-mapping workflow for the interpretable discovery of phase-stable lead-free $A_2BB'X_6$double perovskites. Instead of using machine learning only to rank compositions, the workflow starts from device-relevant targets and maps them back to chemically meaningful descriptor families. By combining charge-balanced enumeration, halide-aware formability filtering, six-family chemical-genome descriptors, ML surrogate models, SHAP interpretation, and DFT validation, the approach connects candidate selection with clear structure–property design rules.

Starting from 13,088 Pb-free charge-balanced compositions, the screening funnel narrowed the search space through stability classification, geometric formability screening, band-gap targeting, and DFT phenotype closure. The final retained set contains seven DFT-validated Pb-free double perovskites: $K_2BePdF_6$, $K_2MnCdCl_6$, $Rb_2TeCuBr_6$, $Cs_2SnGeBr_6$, $Cs_2GeSrBr_6$, $Cs_2NiBaI_6$, and $Cs_2AgInCl_6$. This 99.95% reduction demonstrates that the useful stability–function region is extremely narrow within the accessible $A_2BB'X_6$ chemical space, and that a staged, physically guided workflow is essential for identifying realistic candidates.

The DFT results further show that the validated compounds should not be treated as one generic class. $Rb_2TeCuBr_6$ and $Cs_2SnGeBr_6$ exhibit the most balanced absorber-oriented profiles, combining suitable band gaps, strong optical response, useful dielectric screening, and comparatively favorable transport descriptors. In contrast, $Cs_2GeSrBr_6$ and $Cs_2AgInCl_6$ are better suited to UV-active, transparent, or electron-selective applications, while $K_2BePdF_6$, $K_2MnCdCl_6$, and $Cs_2NiBaI_6$ show more specialized trade-offs.

The principal design outcome is the six-family chemical-genome rule set. Geometric packing genes define the formability window; framework cohesion genes support phase accessibility; polarity and covalency genes tune band-edge hybridization and optical-transition strength; charge-transfer genes regulate chemical plausibility and band alignment; polarization and screening genes control dielectric response; and electronic-identity genes determine band dispersion and carrier asymmetry. These coupled rules demonstrate that band gap alone is insufficient for selecting lead-free optoelectronic materials.

Overall, this work advances lead-free double-perovskite screening from broad forward enumeration toward an interpretable inverse-design strategy. The framework identifies promising candidates. It explains why they survive the screening funnel. It also assigns them to realistic device roles. Future work should extend the approach. This should include finite-temperature stability, defect tolerance, excitonic effects, and spin–orbit coupling where relevant. Experimental synthesis with phase verification should also be considered. Even with these remaining steps, the present workflow provides a reusable target-to-chemistry pathway for discovering Pb-free double perovskites with DFT-validated design rules.

References:

[1] M.-I. Jamesh, H. Tong, M. Du, W. Niu, G. Jia, K.-C. Cheng, C.-W. Hsieh, H.-H. Shen, B. Xu, Y. Tian, X. Xu, H.-Y. Hsu, Advancement of technology towards developing perovskite-based solar cells for renewable energy harvesting and energy transformation applications, Npj Mater. Sustain. 3 (2025) 29. https://doi.org/10.1038/s44296-025-00073-9.

[2] L. Chu, W. Ahmad, W. Liu, J. Yang, R. Zhang, Y. Sun, J. Yang, X. Li, Lead-Free Halide Double Perovskite Materials: A New Superstar Toward Green and Stable Optoelectronic Applications, Nano-Micro Lett. 11 (2019) 16. https://doi.org/10.1007/s40820-019-0244-6.

[3] S. Wang, H. Li, L. Qi, K. Pan, Lead-free halide double-perovskite nanocrystals: structure, synthesis, optoelectronic properties, and applications, J. Mater. Chem. C 13 (2025) 19080–19105. https://doi.org/10.1039/D5TC02430G.

[4] M.H. Moklis, C. Avian, E. Kolor, Md. Rubel, J.S. Cross, Review on Recent Development of Artificial Intelligence and Machine Learning Approaches in Energy Applications, in: M. Elsisi, M. Amer, N. Rinanto, C.-L. Su (Eds.), Adv. Smart Energy Syst. Model. Simul. Secur. Electr. Veh. Microgrids, Springer Nature Switzerland, Cham, 2026: pp. 221–291. https://doi.org/10.1007/978-3-032-17462-8_12.

[5] J.-S. Kim, J. Noh, J. Im, Machine learning-enabled chemical space exploration of all-inorganic perovskites for photovoltaics, Npj Comput. Mater. 10 (2024) 97. https://doi.org/10.1038/s41524-024-01270-1.

[6] E. Landini, K. Reuter, H. Oberhofer, Machine-learning Based Screening of Lead-free Halide Double Perovskites for Photovoltaic Applications, (2022). https://doi.org/10.48550/arXiv.2208.12736.

[7] Z. Chen, J. Wang, C. Li, B. Liu, D. Luo, Y. Min, N. Fu, Q. Xue, Highly versatile and accurate machine learning methods for predicting perovskite properties, J. Mater. Chem. C 12 (2024) 15444–15453. https://doi.org/10.1039/D4TC02268H.

[8] Z. Gao, G. Mao, S. Chen, Y. Bai, P. Gao, C. Wu, I.D. Gates, W. Yang, X. Ding, J. Yao, High throughput screening of promising lead-free inorganic halide double perovskites via first-principles calculations, Phys. Chem. Chem. Phys. 24 (2022) 3460–3469. https://doi.org/10.1039/D1CP04976C.

[9] M.H. Moklis, C. Avian, C. Shuo, S. Boonyubol, J.S. Cross, Machine learning-driven prediction and optimization of selective glycerol electrocatalytic reduction into propanediols, J. Electroanal. Chem. 988 (2025) 119150. https://doi.org/10.1016/j.jelechem.2025.119150.

[10] K. Hippalgaonkar, Q. Li, X. Wang, J.W. Fisher, J. Kirkpatrick, T. Buonassisi, Knowledge-integrated machine learning for materials: lessons from gameplaying and robotics, Nat. Rev. Mater. 8 (2023) 241–260. https://doi.org/10.1038/s41578-022-00513-1.

[11] Y. Wei, J. He, C. Yang, W. Yu, J. Feng, X.-J. Liu, X. Chong, Accelerated Multi-Property Screening of Lead-Free Halide Double Perovskite via Transfer Learning, (n.d.). https://doi.org/10.1002/adfm.202514377.

[12] Z. Guo, B. Lin, Machine learning stability and band gap of lead-free halide double perovskite materials for perovskite solar cells, Sol. Energy 228 (2021) 689–699. https://doi.org/10.1016/j.solener.2021.09.030.

[13] J. Dean, M. Scheffler, T.A.R. Purcell, S.V. Barabash, R. Bhowmik, T. Bazhirov, Interpretable machine learning for materials design, J. Mater. Res. 38 (2023) 4477–4496. https://doi.org/10.1557/s43578-023-01164-w.

[14] J. Riebesell, R.E.A. Goodall, P. Benner, Y. Chiang, B. Deng, G. Ceder, M. Asta, A.A. Lee, A. Jain, K.A. Persson, A framework to evaluate machine learning crystal stability predictions, Nat. Mach. Intell. 7 (2025) 836–847. https://doi.org/10.1038/s42256-025-01055-1.

[15] M. Fronzi, M.J. Ford, K.S. Nayal, O. Isayev, C. Stampfl, Interpretable machine learning for thermoelectric materials design with Kolmogorov–Arnold networks, Sci. Rep. 16 (2026) 14146. https://doi.org/10.1038/s41598-026-44723-x.

[16] J. Udabe, A scientist's guide to AI-driven molecular discovery, Artif. Intell. Chem. 4 (2026) 100107. https://doi.org/10.1016/j.aichem.2026.100107.

[17] H. Wang, R. Ouyang, W. Chen, A. Pasquarello, High-Quality Data Enabling Universality of Band Gap Descriptor and Discovery of Photovoltaic Perovskites, J. Am. Chem. Soc. 146 (2024) 17636–17645. https://doi.org/10.1021/jacs.4c03507.
[18] C.J. Bartel, C. Sutton, B.R. Goldsmith, R. Ouyang, C.B. Musgrave, L.M. Ghiringhelli, M. Scheffler, New tolerance factor to predict the stability of perovskite oxides and halides, Sci. Adv. 5 (2019) eaav0693. https://doi.org/10.1126/sciadv.aav0693.
[19] T. Sato, S. Takagi, S. Deledda, B.C. Hauback, S. Orimo, Extending the applicability of the Goldschmidt tolerance factor to arbitrary ionic compounds, Sci. Rep. 6 (2016) 23592. https://doi.org/10.1038/srep23592.
[20] A. Talapatra, B.P. Uberuaga, C.R. Stanek, G. Pilania, Band gap predictions of double perovskite oxides using machine learning, Commun. Mater. 4 (2023) 46. https://doi.org/10.1038/s43246-023-00373-4.
[21] J. Schmidt, J. Shi, P. Borlido, L. Chen, S. Botti, M.A.L. Marques, Predicting the Thermodynamic Stability of Solids Combining Density Functional Theory and Machine Learning, Chem. Mater. 29 (2017) 5090–5103. https://doi.org/10.1021/acs.chemmater.7b00156.
[22] M. Baharfar, A.C. Hillier, G. Mao, Charge-Transfer Complexes: Fundamentals and Advances in Catalysis, Sensing, and Optoelectronic Applications, Adv. Mater. 36 (2024) 2406083. https://doi.org/10.1002/adma.202406083.
[23] S. Iseki, K. Nonomura, S. Kishida, D. Ogata, J. Yuasa, Zinc-Ion-Stabilized Charge-Transfer Interactions Drive Self-Complementary or Complementary Molecular Recognition, J. Am. Chem. Soc. 142 (2020) 15842–15851. https://doi.org/10.1021/jacs.0c05940.
[24] C. Jelsch, Y. Bibila Mayaya Bisseyou, Deciphering the driving forces in crystal packing by analysis of electrostatic energies and contact enrichment ratios, IUCrJ 10 (2023) 557–567. https://doi.org/10.1107/S2052252523005675.
[25] F. Marin, A. Zappi, D. Melucci, L. Maini, Self-organizing maps as a data-driven approach to elucidate the packing motifs of perylene diimide derivatives, Mol. Syst. Des. Eng. 8 (2023) 500–515. https://doi.org/10.1039/D2ME00240J.
[26] K.M. Steed, J.W. Steed, Packing Problems: High $Z'$ Crystal Structures and Their Relationship to Cocrystals, Inclusion Compounds, and Polymorphism, Chem. Rev. 115 (2015) 2895–2933. https://doi.org/10.1021/cr500564z.
[27] S. Tretiakov, A. Nigam, R. Pollice, Studying Noncovalent Interactions in Molecular Systems with Machine Learning, Chem. Rev. 125 (2025) 5776–5829. https://doi.org/10.1021/acs.chemrev.4c00893.
[28] X. Zhao, M.L. Ball, A. Kakekhani, T. Liu, A.M. Rappe, Y.-L. Loo, A charge transfer framework that describes supramolecular interactions governing structure and properties of 2D perovskites, Nat. Commun. 13 (2022) 3970. https://doi.org/10.1038/s41467-022-31567-y.
[29] F. Gou, Z. Ma, Q. Yang, H. Du, Y. Li, Q. Zhang, W. You, Y. Chen, Z. Du, J. Yang, N. He, J. Luo, Z. Liu, Z. Tian, M. Mao, K. Liu, J. Yu, A. Zhang, F. Min, K. Sun, N. Xuan, Machine Learning-Assisted Prediction and Control of Bandgap for Organic–Inorganic Metal Halide Perovskites, ACS Appl. Mater. Interfaces 17 (2025) 18383–18393. https://doi.org/10.1021/acsami.5c00218.
[30] X. He, J. Liu, C. Yang, G. Jiang, Predicting thermodynamic stability of magnesium alloys in machine learning, Comput. Mater. Sci. 223 (2023) 112111. https://doi.org/10.1016/j.commatsci.2023.112111.
[31] M.R. Soltanian, A. Bemani, F. Moeini, R. Ershadnia, Z. Yang, Z. Du, H. Yin, Z. Dai, Data driven simulations for accurately predicting thermodynamic properties of H2 during geological storage, Fuel 362 (2024) 130768. https://doi.org/10.1016/j.fuel.2023.130768.
[32] H. Zou, H. Zhao, M. Lu, J. Wang, Z. Deng, J. Wang, Predicting thermodynamic stability of inorganic compounds using ensemble machine learning based on electron configuration, Nat. Commun. 16 (2025) 203. https://doi.org/10.1038/s41467-024-55525-y.
[33] C.G. Vera de la Garza, S. Fomine, Machine-learning-accelerated band gap prediction from chemical composition with near-experimental accuracy, Mater. 11 (2026) 101728. https://doi.org/10.1016/j.nxmate.2026.101728.

[34] W. Wang, A. Tudi, R. An, Z. Yang, Interpretable Machine Learning for Bandgap Prediction and Descriptor-Guided Design Rules of Phosphates, Adv. Intell. Discov. n/a (2026) e202600010. https://doi.org/10.1002/aidi.202600010.
[35] Y. Zhuo, A. Mansouri Tehrani, J. Brgoch, Predicting the Band Gaps of Inorganic Solids by Machine Learning, J. Phys. Chem. Lett. 9 (2018) 1668–1673. https://doi.org/10.1021/acs.jpclett.8b00124.
[36] H. Wang, R. Ouyang, W. Chen, A. Pasquarello, High-Quality Data Enabling Universality of Band Gap Descriptor and Discovery of Photovoltaic Perovskites, J Am Chem Soc (2024).
[37] R. Rafiu, M. Sakib Hasan, M. Azizur Rahman, I. Ahamed Apon, K. Kriaa, M. Benghanem, S. AlFaify, N. Elboughdiri, First-principles calculations to investigate structural, electronic, optical, elastic, mechanical and phonon properties of novel Q 3 GaBr 6 (Q = Na and K) for next-generation lead-free solar cells, RSC Adv. 16 (2026) 7803–7829. https://doi.org/10.1039/D5RA10011A.
[38] Y. Zhydachevskyy, Y. Hizhnyi, S.G. Nedilko, I. Kudryavtseva, V. Pankratov, V. Stasiv, L. Vasylechko, D. Sugak, A. Lushchik, M. Berkowski, A. Suchocki, N. Klyui, Band Gap Engineering and Trap Depths of Intrinsic Point Defects in RAlO3 (R = Y, La, Gd, Yb, Lu) Perovskites, J. Phys. Chem. C 125 (2021) 26698–26710. https://doi.org/10.1021/acs.jpcc.1c06573.
[39] M.U. Ghani, M. Junaid, K.M. Batoo, M.F. Ijaz, B. Zazoum, An extensive study of structural, electronic, optical, mechanical, and thermodynamic properties of inorganic oxide perovskite materials ScXO3 (X = Ga, In) for optoelectronic applications: A DFT study, Inorg. Chem. Commun. 172 (2025) 113459. https://doi.org/10.1016/j.inoche.2024.113459.
[40] Y.K. Chung, J. Lee, W.-G. Lee, D. Sung, S. Chae, S. Oh, K.H. Choi, B.J. Kim, J.-Y. Choi, J. Huh, Theoretical Study of Anisotropic Carrier Mobility for Two-Dimensional Nb2Se9 Material, ACS Omega 6 (2021) 26782. https://doi.org/10.1021/acsomega.1c03728.
[41] Y.K. Chung, J. Lee, W.-G. Lee, D. Sung, S. Chae, S. Oh, K.H. Choi, B.J. Kim, J.-Y. Choi, J. Huh, Theoretical Study of Anisotropic Carrier Mobility for Two-Dimensional Nb2Se9 Material, ACS Omega 6 (2021) 26782–26790. https://doi.org/10.1021/acsomega.1c03728.
[42] J. Laflamme Janssen, Y. Gillet, S. Poncé, A. Martin, M. Torrent, X. Gonze, Precise effective masses from density functional perturbation theory, Phys. Rev. B 93 (2016) 205147. https://doi.org/10.1103/PhysRevB.93.205147.
[43] Z. Li, P. Graziosi, N. Neophytou, Deformation potential extraction and computationally efficient mobility calculations in silicon from first principles, Phys. Rev. B 104 (2021) 195201. https://doi.org/10.1103/PhysRevB.104.195201.
[44] F. Murphy-Armando, G. Fagas, J.C. Greer, Deformation Potentials and Electron−Phonon Coupling in Silicon Nanowires, Nano Lett. 10 (2010) 869–873. https://doi.org/10.1021/nl9034384.
[45] C. Chen, Y. Zuo, W. Ye, X. Li, Z. Deng, S.P. Ong, A Critical Review of Machine Learning of Energy Materials, Adv. Energy Mater. 10 (2020) 1903242. https://doi.org/10.1002/aenm.201903242.
[46] A. Mazheika, Y.-G. Wang, R. Valero, F. Viñes, F. Illas, L.M. Ghiringhelli, S.V. Levchenko, M. Scheffler, Artificial-intelligence-driven discovery of catalyst genes with application to CO2 activation on semiconductor oxides, Nat. Commun. 13 (2022) 419. https://doi.org/10.1038/s41467-022-28042-z.
[47] Y. Zhang, Y. Xia, A. Shakiba, H. Zhang, X. Hao, P.V. Kumar, M.P. Suryawanshi, Machine Learning for Designing Perovskites and Perovskite-Inspired Solar Materials: Emerging Opportunities and Challenges, Adv. Sci. 13 (2026) e74952. https://doi.org/10.1002/advs.74952.